\documentclass[varenna]{cimento}
\usepackage{graphicx} 
\usepackage[utf8]{inputenc}
\usepackage[normalem]{ulem} 
\usepackage{xcolor}
\usepackage{bm}
\usepackage{amssymb}
\usepackage[english]{babel}
\usepackage{array}
\usepackage{cite}

\title{
Mass-imbalanced Fermi mixtures with resonant interactions
}
\author{M.~Zaccanti\thanks{zaccanti@lens.unifi.it}
}
	\institute{European Laboratory for Nonlinear Spectroscopy (LENS), University of Florence, 50019 Sesto Fiorentino, Italy\\
Istituto Nazionale di Ottica del Consiglio Nazionale delle Ricerche (CNR-INO) c/o LENS, 50019 Sesto Fiorentino, Italy}

\begin{document}

\maketitle

\begin{abstract}
In these notes I provide an overview of ongoing theoretical and experimental research on ultracold atomic mixtures composed by two different fermionic species. 
First, I describe a general and simple framework that should allow also a non-expert reader to understand the rich few-body phenomena connected with such systems, and their possible impact at the many-body level.
I then move to discuss the specific combination of fermionic lithium ($^6$Li) and chromium ($^{53}$Cr) atoms, currently investigated in our lab, highlighting its peculiar properties with respect to other Fermi mixtures nowadays available. 
Finally, I summarize recent experimental progress achieved in producing and characterizing this novel  system,  providing an outlook for future studies based on ultracold $^6$Li-$^{53}$Cr Fermi mixtures. 
\end{abstract}

\section{Introduction} \label{Introduction}
Quantum matter composed by unequal kinds of fermionic particles, such as quarks of different colors, or electrons belonging to different lattice bands, is known to promote a plethora of exotic phenomena \cite{Casabuoni2004,Andersen2011,Hammer2013,Yi2017,Wang2018,Jiang2021}, qualitatively distinct from those characterizing single-component systems. 
The combination of quantum statistics with a mass asymmetry and a distinct response to external fields of two different fermionic species, indeed provides an increased level of complexity, with a strong impact both at the few- and many-body level.
In this context, heteronuclear Fermi mixtures of ultracold atoms, resonantly interacting close to a Feshbach resonance \cite{Chin2010}, are regarded as clean and versatile frameworks optimally-suited for 
the disclosure of exotic few-particle cluster states \cite{Efimov1970,Efimov1973,Kartavtsev2007,Nishida2008,Levinsen2009,Castin2010,Endo2011,Blume2012,Endo2012,Ngampruetikorn2013,Levinsen2013,Bazak2017,Bazak2017b, Liu2022A}, 
the investigation of novel types of impurity problems \cite{Mathy2011,Massignan2014,Schmidt2018,Liu2022B,Scazza2022} and mediated quasi-particle interactions \cite{Bulgac2006,Suchet2017,Camacho-Guardian2018}, 
and the exploration of elusive many-body regimes -- primarily in the context of unconventional superfluid pairing \cite{Liu2003,Forbes2005,Iskin2006,Parish2007,Baranov2008,Gubbels2009,Gezerlis2009,Baarsma2010,Gubbels2013,Wang2017,Pini2021} beyond the BCS-BEC crossover scenario \cite{Varenna2007,Giorgini2008,Zwerger2012}, and of quantum magnetism \cite{Keyserlingk2011,Sotnikov2012,Sotnikov2013,Massignan2013,Cui2013}.
Additionally, weakly-bound bosonic dimers created from a Fermi mixture, thanks to their increased collisional stability near an $s$-wave Feshbach resonance \cite{Petrov2004,Jag2016}, represent an unparalleled starting point to realize degenerate samples of ground-state polar molecules \cite{Ni2008,Takekoshi2014,Molony2014,Park2015,Guo2016,Son2020}.

In particular, Fermi mixtures with mass ratios 8.17$\leq M/m\!\leq$13.6 are especially appealing from a few-particle physics perspective, as they are predicted to support, already in three dimensions, non-Efimovian  cluster states, experimentally unexplored thus far, which exhibit universal character and a peculiar $p$-wave orbital symmetry \cite{Kartavtsev2007,Endo2011,Endo2012,Blume2012A,Bazak2017}. 
These elusive few-body states are extremely relevant also from a many-body viewpoint, in light of their collisional stability: Owing to the \textit{halo} nature of such non-Efimovian clusters, largely exceeding in size the van der Waals range of the interatomic interaction potential, their existence indeed does not trigger an increase of inelastic decay processes \cite{Levinsen2011}, in stark contrast to the widely-explored Efimovian case \cite{Naidon2017}. Therefore, the presence of fermionic trimers \cite{Kartavtsev2007,Endo2011,Endo2012} and bosonic tetramers \cite{Blume2012A,Bazak2017} -- expected to exist for 8.17$\leq M/m\!\leq$13.6 \cite{Kartavtsev2007} and 8.86$\leq M/m\!\leq$13.6 \cite{Bazak2017}, respectively -- may uniquely allow one to experimentally attain qualitatively new many-body regimes, within which strong few-body correlations add to, or may even overcome, the standard two-body ones.

Among other Fermi-Fermi mixtures nowadays available, i.e. $^6$Li-$^{40}$K \cite{Wille2008,Voigt2009,Costa2010}, $^{40}$K-$^{161}$Dy \cite{Ravensbergen2018,Ravensbergen2020}, $^{6}$Li-$^{173}$Yb \cite{Hara2011,Green2020} and $^6$Li-$^{167}$Er \cite{Schaefer2023}, here I will focus on the special case of a degenerate Fermi mixture made of $^6$Li alkali and $^{53}$Cr transition metal ultracold atoms, recently realized, and uniquely available, in our lab in Florence \cite{Neri2020,Ciamei2022,Ciamei2022B}.
Our interest for this new system is three-fold: 
First, the peculiar chromium-lithium mass ratio, of about 8.8, is extremely close to the critical values above which both three- and four-body non-Efimovian cluster states are predicted to emerge \cite{Kartavtsev2007,Endo2011,Endo2012,Blume2012A,Bazak2017}.
This feature, combined with the recent discovery of various magnetic Feshbach resonances well suited to control Li-Cr interactions \cite{Ciamei2022B}, makes such a bi-atomic combination an unparalleled framework with which to explore a new class of \textit{elastic} few-body effects, and their impact at the many-body level \cite{Endo2012,Bazak2017}.
Second, in the regime of strong repulsive interactions, three-body recombination processes are predicted to be drastically suppressed for the specific Li-Cr mass ratio \cite{Petrov2003}, lithium-chromium Fermi mixtures thus representing a pristine platform to explore Stoner's ferromagnetism \cite{Stoner1933} and related phenomena \cite{Jo2009,Sanner2009,Scazza2017,Valtolina2017,Amico2018,Scazza2020}, `immune' to the pairing instability. 
Finally, recent \textit{ab initio} calculations \cite{Zaremba2022} foresee, for the ground state of the LiCr dimer, a sizable electric dipole moment of about 3.3 Debye, combined with a $S\!=\!5/2$ electronic spin, thereby making Li-Cr mixtures also extremely appealing candidates to realize ultracold paramagnetic polar molecules.

These notes are organized as it follows: In Section~\ref{Sec2and3} I provide a simple theoretical framework --  summarizing the more detailed discussion presented in Refs.~\cite{Levinsen2011,Petrov2012} -- that should enable also a non-expert reader to understand, and appreciate, the rich phenomenology that characterizes few-fermion systems under strong interactions. 
In Section~\ref{SecLiCrTheo} I focus on the specific $^6$Li-$^{53}$Cr Fermi mixture we are currently investigating in our lab, discussing the peculiar properties of such a novel system. 
Finally, in Section~\ref{SecLiCrExp} I summarize the most recent progress in making, probing and understanding this novel bi-atomic mixture in experiments, concluding with an outlook for future studies.  
  
\section{Basic properties of two- and three-body fermion systems} \label{Sec2and3}
In this section, I provide an overview of the physical phenomena and theoretical tools relevant to understand few-fermion systems in the ultracold regime.
I first recall some textbook results on two-particle scattering, with special emphasis on low-energy collisions. The interested reader can find complete derivation and a more thorough discussion of the results reported herein in Refs.~\cite{Chin2010, Landau1977book, Petrov2012}.
I then provide an intuitively simple description of three-fermion systems based on the Born-Oppenheimer approximation valid for $M/m \gg 1$ \cite{B-O_App1927,Fonseca1979}, and discuss the impact of the mass asymmetry and of the specific nature of the two-body interaction on the resulting few-body phenomena. The discussion closely follows the one of Refs.~\cite{Levinsen2011,Petrov2012}, to which I refer the  reader for further details. 

\subsection{Two-body scattering and magnetic Feshbach resonances} \label{2B}

The collision between two quantum particles of mass $M$ and $m$, interacting via a potential $V(\textbf{r})$ -- with $\textbf{r}$ denoting the inter-particle separation -- modifies the properties, and correspondingly the wavefunction, of the two-body system.
A non-zero $V(\textbf{r})$ changes the relative motion of the pair that is otherwise described, in the center-of-mass reference frame, by a plane wave $\sim e^{i k z}$, characterized by relative momentum $\textbf{k}$ and energy $\hbar^2 k^2/(2 m_r)$, $m_r=M m/(M+m)$ being the reduced mass.
Specifically, a finite interaction gives rise to an additional contribution to the system wavefunction which, for $r \rightarrow \infty$, has the asymptotic form of an outgoing spherical wave $\sim e^{ik'r}/r$ which adds to the incoming plane wave. 
The amplitude of such a diffused wave, the \textit{scattering amplitude}, typically denoted by $f(\textbf{k},\textbf{k'})$, encodes all information about the collision process, and it generally depends upon the incoming (outgoing) momentum $\textbf{k}$ ($\textbf{k'}$).

Focusing on \textit{elastic} scattering processes such that $|\textbf{k}|=|\textbf{k'}|=k$, and restricting to the case of central potentials of the kind $V(\textbf{r})=V(r)$, the scattering amplitude can be conveniently decomposed into a sum over partial-wave channels with defined angular momentum quantum number $l$ \cite{Landau1977book}: 
\begin{equation}
 \label{ftot}
 f(\textbf{k},\textbf{k'})  =  f(k, \theta)  =  \sum_{l=0}^{\infty} (2l+1) P_l(cos\theta) f_l(k).
\end{equation}
Here $P_l$ is the $l^{th}$ Legendre polynomial and $\theta=\angle_{\textbf{k},\textbf{k'}}$ the scattering angle.
In the case of two indistinguishable particles, Eq.~(\ref{ftot}) must be modified to take into account the symmetry (antisymmetry) properties of the pair of bosons (fermions) under particle exchange: Given that the Legendre polynomials have $(-1)^{l}$ parity, the partial wave expansion for identical bosons (fermions) will contain only even (odd) waves\footnote{Moreover, a factor of 2 appears in Eq.~(\ref{ftot}) for identical particles, arising from the (anti)-symmetrization of the wavefunction.}. 
Each $f_l(k)$ can be conveniently expressed in terms of an associated phase shift $\delta_l(k)$ in the following equivalent forms \cite{Landau1977book} 
\begin{equation}
 \label{fl.all}
 f_l(k) = \frac{1}{2 i k}(e^{2 i \delta_l(k)}-1)
 =  \frac{1}{k cot \delta_l(k)-i k}             
 = \frac{sin(2\delta_l(k))}{2k}+i \frac{sin^2(\delta_l(k))}{k}.
\end{equation}
Solving the scattering problem for a given $V(r)$ thus amounts to gain full knowledge of all corresponding $\delta_l(k)$ functions. 

From the scattering amplitude one obtains the scattering cross section upon integration over the solid angle: 
\begin{equation}
 \label{sigma}
 \sigma(k) = \int |f(k, \theta)|^2 d\Omega = 4 \pi \sum_{l=0}^{\infty} (2l+1) \frac{sin^2(\delta_l(k))}{k^2}\equiv \sum_{l=0}^{\infty} \sigma_l(k). 
\end{equation} 
By comparing Eqns.~(\ref{sigma}) and (\ref{fl.all}), and recalling that $P_l(1)=1$ for each $l$, it is straightforward to verify that
\begin{equation}
 \label{OpTheo}
 \sigma(k) = \frac{4\pi}{k} Im f(k,0).  
\end{equation}  
In particular, one can notice how each partial-wave component in Eq.~(\ref{sigma}) reaches its maximum value,  the so-called \textit{unitary limit},   
\begin{equation}
 \label{sigma_lMAX}
 \sigma_{l,MAX}(k) = (2l+1) \frac{4 \pi}{k^2} \qquad
                     \tx{when\ } \delta_l(k)=\frac{\pi}{2}.
\end{equation}
Similarly, from the knowledge of $f(k, \theta)$ one obtains the \textit{elastic scattering rate} $1/\tau$. 
The latter, which generally depends on the density of colliding partners, can be suitably evaluated within the framework of the so-called impact theory of pressure broadening \cite{Baranger1958a,Baranger1958b}, as long as the collision of a particle within a medium (at density $\bar{n}$) can be considered as being effectively instantaneous. 
Under this assumption, the scattering rate is linked to the \textit{forward} scattering amplitude averaged over all collision momenta, $\left\langle f(k, \theta=0) \right\rangle$, through the useful relation
\begin{equation}
 \label{scatrate}
 1/\tau = \bar{n} \, \frac{\hbar}{m_r} \,4 \pi \, Im\left\langle f(k, 0)\right\rangle\!. 
\end{equation} 
Additionally, $f(k, \theta)$ provides information also about the \textit{energy shift} $h \delta \nu$, experienced by one particle due to its interaction with the surrounding medium. 
Again within the framework of the impact theory \cite{Baranger1958a,Baranger1958b}, one finds that
\begin{equation}
 \label{enshift}
\delta \nu = - \bar{n} \, \frac{\hbar}{m_r} \,Re\left\langle f(k, 0)\right\rangle\!. 
\end{equation}
As a general result, a positive (negative) value of $Re\left\langle f(k, 0)\right\rangle$ corresponds to a net attractive (repulsive) interaction energy.
  
A great simplification to the scattering problem occurs if we focus on slow collisions ($k \rightarrow$0) off interaction potentials that are \textit{short-ranged}: I.e., potentials that drop like $V(r)\sim r^{- \alpha}$ with $\alpha\!>3$, for $r \rightarrow \infty$. 
This case is indeed relevant when we consider pai\!rs of (non-magnetic) atoms, which feature a van der Waals interaction of the kind $V_{vdW}(r)\! = -C_6/r^6\!= -E_{vdW}(r/R_e)^6$, characterized by the range $R_e\!=(\frac{2 m_r C_6}{\hbar^2})^{1/4}$ and energy $E_{vdW}\!=\frac{\hbar^2}{2 m_r R_e^2}$, respectively \cite{Chin2010}.
For short-ranged $V(r)$, it can be shown \cite{Landau1977book} that, for each $l$ value, 
\begin{equation}
 \label{dl}
 \delta_l(k) \sim k^{2l+1} \qquad
                     \tx{for\ } k \rightarrow 0
\end{equation}
and hence, from Eq.~(\ref{fl.all}), that $f_l(k) \sim k^{2l}$.
As a consequence, for atoms colliding in the ultracold regime, where the De Broglie wavelength greatly exceeds the van der Waals range ($k R_e\!\ll 1$), the resulting scattering is predominantly isotropic -- i.e. $s$-wave -- as for a light wave incident on an object that is much smaller than its wavelength.

In particular (see e.g. Ref.~\cite{Petrov2012} for details), the net effect of $V(r)$ on the two-body system at large inter-particle separation, $r\!\gg R_e$, can be formally taken into account by imposing a boundary condition at the origin for the log-derivative of the radial wavefunction $\psi(\textbf{r})$:
\begin{equation}
 \label{BP}
\frac{[r\psi]'}{[r\psi]}|_{r \rightarrow 0} =k \,cot \delta_0(k),
\end{equation}
the so-called Bethe-Peierls boundary conditions \cite{Bethe1935}. 
While in the generic scenario the replacement of $V(r)$ by the condition Eq.~(\ref{BP}) does not simplify the problem -- as it requires to know the phase shift, anyway -- this becomes a valuable tool in the case of low-energy scattering, $k R_e\! \ll 1$.
In this regime, indeed, Eq.~(\ref{fl.all}) can be expanded in powers of the small parameter $k R_e$ resulting, for the s-wave channel ($l=0$), in a trend like \cite{Landau1977book}
\begin{equation}
 \label{Rs_exp}
k\, cot \delta_0(k) \approx -\frac{1}{a} - R^*k^2 + ...,
\end{equation}
where the constants $a$ and $R^*$ are defined as the \textit{scattering length} and the \textit{effective range parameter}, respectively.  
Correspondingly, the $s$-wave scattering amplitude becomes:
\begin{equation}
 \label{f0}
f_0(k) = -\frac{1}{\frac{1}{a} + R^*k^2 + ik}\!.
\end{equation}
The scattering problem in such a regime can thus be solved by imposing the boundary condition Eq.~(\ref{BP}) --  now solely depending on two $k$-independent parameters that may be eventually  determined experimentally, while yet ignoring the detailed form of the true interaction potential. The latter information is indeed not necessary in this case: All short-range potentials are equivalent, as long as they feature the same $a$ and $R^*$, so that one can choose an idealized zero-range (pseudo-)potential to reproduce Eq.~(\ref{f0}). 

One important feature of Eq.~(\ref{f0}) is that it exhibits the well-known Breit-Wigner resonance shape \cite{Landau1977book} -- relevant to describe low-energy collisions where the scattering state, at energy $E=\hbar^2 k^2/(2 m_r)$, is coupled to a quasi-stationary state of energy $E_{res}$, through a \textit{coupling amplitude} $\gamma$ -- for which:
\begin{equation}
 \label{fBW}
f_{BW}(E) = -\frac{\hbar \gamma/\sqrt{2 m_r}}{E-E_{res} + i\gamma \sqrt{E}}.
\end{equation}
By comparing Eq.~(\ref{fBW}) with Eq.~(\ref{f0}), equivalently recast in energy units, it is easy to verify the link between the parameters ($a$, $R^*$) and ($E_{res}$,$\gamma$), respectively:
\begin{eqnletter}
 \label{f0vsfBW}
 a & = & - \frac{\hbar \gamma}{\sqrt{2 m_r} E_{res}}   \label{aBW}\\
 R^* & = &  \frac{\hbar}{\sqrt{2 m_r} \gamma}       \label{RsBW}.
\end{eqnletter}
A few important remarks are due here. 
First, while $a$ depends on both $\gamma$ and $E_{res}$, $R^*$ is only determined by the coupling amplitude. In equivalent terms, the coupling \textit{energy} between the scattering and quasi-stationary states equals
\begin{equation}
 \label{gamma}
\gamma^2 = \frac{\hbar^2 }{2 m_r R^{*2}}.
\end{equation}  
Strong (weak) couplings thus correspond to small (large) effective range values in the scattering amplitude, and vice-versa. 
Furthermore, since $2 \sqrt{E} \gamma$ is nothing but the decay rate of the quasi-stationary state \cite{Petrov2012}, both $\gamma$ and $R^*$ are positive-defined quantities.
Second, from Eq.~(\ref{aBW}) one sees how the sign of the scattering length is determined by the one of $E_{res}$: Namely, $a\!>$0 for $E_{res}\!<$0, i.e. the scattering length is positive only if a real bound state exists. 
Vice-versa, $a\!<$0 when the quasi-stationary level is a virtual state at $E_{res}\!>$0. 
Third,  the magnitude of $a$ can be arbitrarily tuned if the energy of the quasi-stationary state can be varied around the scattering threshold. In particular, when $E_{res}\!\rightarrow 0$, $|a|\!\rightarrow \infty$ and, correspondingly, $\delta_0\!\rightarrow \pi/2$, see Eq.~(\ref{Rs_exp}). 
 
While in most physical cases $E_{res}$ is fixed, for several ultracold systems this parameter can be tuned via the Zeeman effect, giving rise to the Feshbach resonance (FR) phenomenon \cite{Chin2010}. 
This occurs when the scattering threshold of two atoms in a certain hyperfine and Zeeman state combination (the \textit{open} channel) is coupled through a non-zero $\gamma$ to a nearly-degenerate molecular state, supported by the interatomic potential asymptotically connected with another hyperfine domain (the \textit{closed} channel). 
Owing to a non-vanishing differential magnetic moment $\delta \mu= \mu_c-\mu_o$ between closed and open channels, one thus obtains in this case $E_{res}=\delta \mu (B-B_0)$, $B$ denoting an external magnetic-field bias $B$, and $B_0$ the field value at which the bound state in the closed channel is degenerate with the scattering threshold.
As a result, near a FR one obtains from Eq.~(\ref{aBW}) that the scattering length is resonantly enhanced with respect to its background value, with a $B$-field dependence of the kind   
\begin{equation}
 \label{aB1}
a(B) = -\frac{\hbar \gamma}{\sqrt{2 m_r} \delta \mu (B-B_0)}.
\end{equation}   
By denoting with $a_{bg}$ the scattering length value far from the resonance center $B_0$, one can define the magnetic width of the FR as:
\begin{equation}
 \label{deltaB}
\Delta B\equiv \frac{\hbar \gamma}{\sqrt{2 m_r} a_{bg}\delta \mu}.
\end{equation} 
Accounting also for a non-zero, off-resonant contribution, Eq.~(\ref{aB1}) can be then recast in the familiar form~\cite{Chin2010}
\begin{equation}
 \label{aB2}  
a(B) = a_{bg}(1-\frac{\Delta B}{B-B_0}).
\end{equation} 
Furthermore, inserting the expression for the coupling amplitude $\gamma$ given by Eq.~(\ref{deltaB}) into Eq.~(\ref{RsBW}), the effective range parameter reads \cite{Petrov2004B} 
\begin{equation}
 \label{RsB}
R^* = \frac{\hbar^2}{2m_{r}}\frac{1}{a_{bg}\Delta B \delta \mu}.
\end{equation}  
By recalling that large (small) $\gamma$ values physically imply strong (weak) coupling between the scattering channel and the quasi-discrete level,  one can equivalently classify a Feshbach resonance as \textit{broad} or \textit{narrow}, depending whether the effective range parameter in Eq.~(\ref{RsB}) is small or large compared with the true range of the potential: $R^*\! \leq R_e \Rightarrow$ broad resonance; $R^*\! \gg R_e  \Rightarrow$ narrow resonance. 
Note that an essentially analogous classification is made in Ref.~\cite{Chin2010} in terms of the `\textit{dimensionless resonance strength parameter}', which can be expressed as $s_{res}\sim 0.96 R^*/R_e$: broad (narrow) FRs are characterized by large (small) $s_{res}$ values. 

Let us now move to consider how the finite coupling with the open channel modifies properties of the bound state near the FR, otherwise represented by a closed-channel molecule for $\gamma=0$. 
To this end, we look for the pole of $f_0(k)$ in Eq.~(\ref{f0}) at negative energies: I.e, we replace $k\rightarrow i \kappa$, with $\kappa>0$, so that $\epsilon_b=-\frac{\hbar^2 \kappa^2}{2 m_r}<0$, with the associated wavefunction falling like $\psi_b(r)\sim e^{-\kappa r}/r$ at $r\gg R_e$. We thus look when
\begin{equation}
 \label{pole}
\frac{1}{a}-R^* \kappa^2-\kappa=0.
\end{equation}
It is easy to verify that this happens only when $a\!>$0, for
\begin{equation}
 \label{astar}
\kappa=\frac{\sqrt{\frac{4 R^*}{a}+1}-1}{2R^*}\equiv \frac{1}{a^*},
\end{equation}
thus yielding
\begin{equation}
 \label{Eb}
\epsilon_b=-\frac{\hbar^2}{2 m_r a^{*2}}.
\end{equation}
The above result interpolates between two familiar forms obtained in the limits of small and large $R^*/a$ values, respectively:
For $R^*/a \ll 1$ -- i.e. close to the resonance pole, or in the broad resonance case -- one has $1/a^*\sim 1/a$, so that Eq.~(\ref{Eb}) becomes 
\begin{equation}
 \label{halo}
\epsilon_b|_{R^*\!/a\! \ll 1} \sim -\frac{\hbar^2}{2 m_r a^{2}} \equiv \epsilon_0.
\end{equation} 
In this case, the bound state energy features a typical parabolic trend $\epsilon_0\propto -1/a^2 \propto -(B-B_0)^2$: the dimer is a halo state whose energy and wavefunction solely depend on the scattering length $a$ (universal regime). 
In the opposite limit $R^*/a \gg 1$, instead, $1/a^*\sim 1/\sqrt{a R^*}$ and, exploiting Eqns.~(\ref{aB2}) and (\ref{RsB}), one finds a dimer energy trend 
\begin{equation}
 \label{ccmol}
\epsilon_b|_{R^*\!/a\! \gg 1} \sim -\frac{\hbar^2}{2 m_r a R^*}=\delta \mu (B-B_0).
\end{equation}      
In this case, the bound state energy linearly decreases with the field detuning, and it coincides with the one of the bare closed-channel molecule.
It is important to emphasize that the dimensionless parameter $R^*/a$ also quantifies the open/closed channel fraction characterizing the (dressed) Feshbach molecule in the zero-range approximation. Referring the interested reader to Ref.~\cite{Petrov2012} for the derivation of the result, it is also useful to keep in mind that the dimer open-channel fraction is given by 
\begin{equation}
 \label{Popen}
P_{open} = \frac{1}{\sqrt{1+4R^*/a}},
\end{equation}
and, correspondingly, the closed-channel fraction is obtained as $P_{closed}=1-P_{open}$. 
It is easy to verify that, for $P_{open}\!\sim$1, the binding energy  follows the universal behavior Eq.~(\ref{halo}) whereas, in the opposite limit $P_{open}\!\ll$1, the Feshbach dimer essentially coincides with the bare closed-channel molecular state,  Eq.~(\ref{ccmol}).
Correspondingly, the magnetic moment associated with the Feshbach dimer is given by $\partial \epsilon_b/\partial B\!=\mu_o P_{open}+\mu_c (1-P_{open})$ at all detunings: As such, experimental measurement of the dimer magnetic moment around a FR provides information about the magnetic-field dependence of the open-channel fraction, thereby of $R^*/a$ -- see Ref.~\cite{Soave2023} and R. Grimm's contribution to these Proceedings. 
In general, the transition from one regime to the other will occur at magnetic field detunings that depend on the character of the resonance considered: For broad resonances, the universal regime Eq.~(\ref{halo}), with $P_{open}\!\sim$1, will extend over a $B$-field region of order $\Delta B$ from the resonance center $B_0$, whereas for narrow FRs this will only occur for detunings $|B-B_0|\!\ll\!\Delta B$.
\begin{figure}[t!]
\centering
\includegraphics[width=.8\columnwidth]{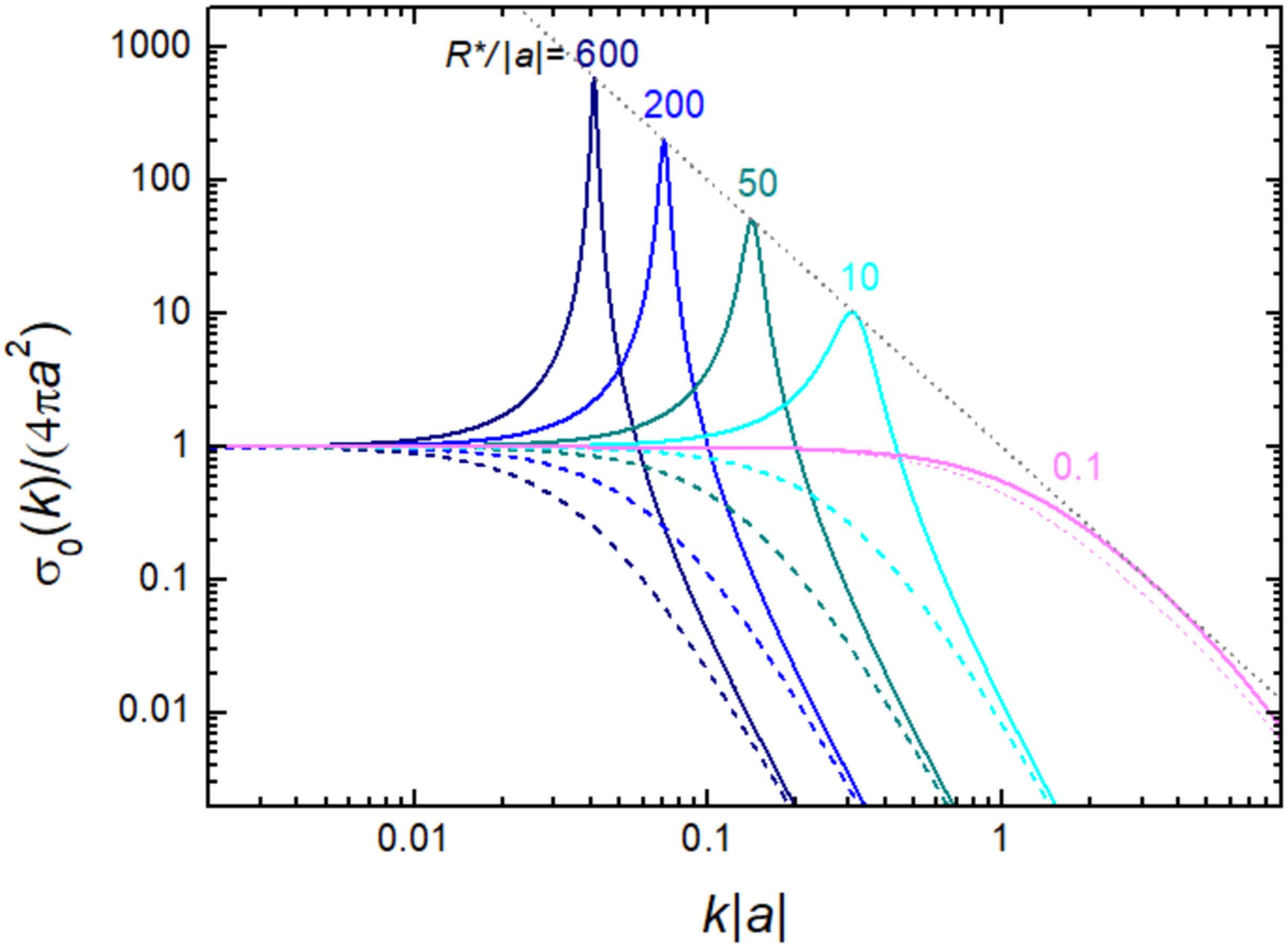} 
\vspace*{-0pt}
\caption{Normalized scattering cross section $\sigma_0(k)/\sigma_0(0)$ as a function of $k|a|$, for different values of $R^*/|a|$, see legend. Solid (dashed) lines refer to negative (positive) $R^*/a$. 
For $a\!>$0 and $R^*/a\!\gg$1, $\sigma_0(k)$ is maximum at $k\!=$0 while laying systematically below the curve corresponding to the broad resonance limit $R^*/|a|\!\ll$1 (see magenta curves). For $a\!<$0, instead, as long as $R^*/|a|\!\geq\!1/2$, the cross section exhibits a sharp, low-momentum peak, centered at $k_{M}$ where $\sigma_0(k)$ reaches its unitary limit $\sigma_{0,MAX}(k)\!=4\pi/k_M^2$, indicated by the dotted gray line.}
\label{sigma_theo}
\vspace*{-20pt}
\end{figure}

At a first glance, the distinction between narrow and broad FRs does not seem to cause a significant change: From a two-body perspective, any broad resonance becomes narrow when the incoming momenta and/or detunings are sufficiently large. Vice-versa, also narrow resonances will behave as broad ones for sufficiently small $k$ and $|B-B_0|$ values, so that $R^*k^2\!\ll\!1/a$ in Eq.~(\ref{f0}). 
In practice, though, such a distinction is highly relevant, given that in realistic situations one will have (i) a finite $B$-field stability, that limits the accuracy with which to access the $R^*\!\ll a$ region in experiments; (ii) a finite momentum distribution, either due to a finite thermal spread at $T\!>$0 or, for fermionic samples, to the presence of a finite Fermi momentum $\kappa_F$ even at zero temperature. In the broad resonance case, one will typically have $R^*\!\sim\!R_e\!\ll 1/\kappa_F$, and thus all momenta $k$ will simultaneously reach unitary conditions $\delta_0(k)\!=\pi/2$ for $1/a\!=$0. In turn, for narrow resonances and realistic densities $n=\kappa_F^3/(6 \pi^2)$, $R^* \kappa_F\! \gg 1$, so that different momenta will reach the unitary limit, see Eq.~(\ref{sigma_lMAX}), at different detunings.
This makes that the low-temperature many-body regimes which can be accessed near broad and narrow resonances are qualitatively different, see e.g. Refs.~\cite{Varenna2007,Giorgini2008,Zwerger2012} and \cite{Radzihovsky2010}.

This important point is better clarified if we look at how the scattering cross-section $\sigma_0(k)$, derived from Eq.~(\ref{f0}), changes with the parameter $R^*/a$, characterizing the `narrowness' of the resonance. To this end, consider the dimensionless form
\begin{equation}
 \label{sigmaNorm}
\frac{\sigma_0(k)}{\sigma_0(0)} =\frac{\sigma_0(k)}{4 \pi a^2} = \frac{1}{(1+(k a)^2\frac{R^*}{a})^2+(ka)^2},
\end{equation}     
 shown in Fig.~\ref{sigma_theo} for different  negative (solid lines) and positive (dashed lines) $R^*/a$ values, respectively.
One can notice the following features:
First, as $k \rightarrow 0$, all curves converge to unity, i.e. to the same $\sigma_0(0)$, which solely depends on the scattering length (`broad resonance limit').
Second, for small but finite $k$, there is a clear asymmetry between the $a\!>$0 and $a\!<$0 cases for $R^*/|a|\! \gg$1, which vanishes instead when $R^*/|a|\!\ll$1 (see magenta lines). 
Indeed, for $a\!<0$, $\sigma_0(k)$ is characterized by a sharp peak, centered at $k_{M}\!=\frac{1}{R^*}\sqrt{\frac{R^*}{|a|}-\frac{1}{2}}$, where the cross-section reaches its unitary limit $\sigma_0(k_{M})\!=4 \pi / k_{M}^2$, a value which may greatly exceed $\sigma_0(0)$. 
Such a peak at $k_M\!>$0, present as long as $R^*/|a|\!\geq\!1/2$ and absent instead for all $a\!>0$ values, is characterized by a full width at half maximum that, for $R^*/|a|\!\gg$1, is given by $\Delta k_{M}\!\sim \frac{1}{R^*}(1+\frac{|a|}{4R^*})$.
Third, one can notice how, at low momenta, the curves with $R^*/|a|\!\gg$1 lay systematically above (below) the ones with $R^*/|a|\!\ll$1.
\begin{figure}
\centering
\includegraphics[width=.8\columnwidth]{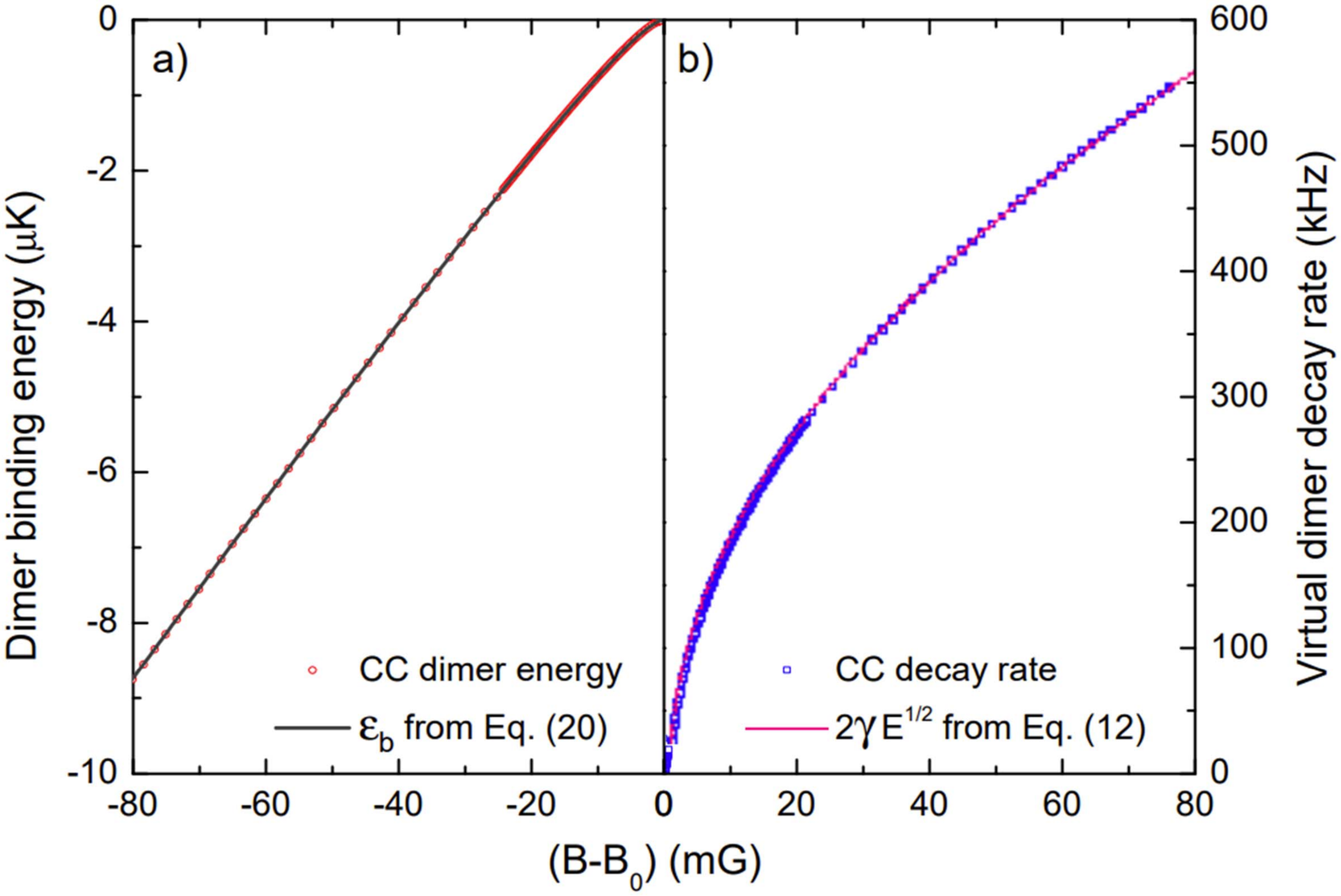} 
\vspace*{-20pt}
\caption{Dimer binding energy (panel \textbf{a}, red circles) and decay rate of the virtual state (panel \textbf{b}, blue squares), obtained from coupled-channel calculations for one Li-Cr FR located around 1461 G \cite{Ciamei2022B}, are compared with the corresponding trends provided by the zero-range approximation discussed in the text. By optimizing the values of $a_{bg}$, $\Delta B$ and $\delta\mu$ -- employed as fitting parameters -- the simple zero-range model is able to reproduce the results of the (much more involved) CC calculations with remarkable accuracy.}  
\label{CCvsZR}
\vspace*{-15pt}
\end{figure}

In conclusion of this part, it is also useful to emphasize how the zero-range approximation summarized above is able to provide a \textit{quantitatively} accurate description of real physical systems. 
As an example, in Fig.~\ref{CCvsZR} I show the dimer binding energy (red circles) and  decay rate of the virtual state (blue squares) near one $s$-wave FR occurring in the lithium-chromium ($^6$Li-$^{53}$Cr) system around 1461 G, obtained from coupled-channel (CC) calculations, optimized by fitting more than 50 FRs experimentally determined \cite{Ciamei2022B}. The CC data are compared with the corresponding trends (solid lines) given by the zero-range approximation -- see Eqs.~(\ref{f0vsfBW})-(\ref{RsB})-- with best-fit values $a_{bg}$=41.48 $a_0$, $\Delta B$=0.477 G, $\delta\mu$=2 $\mu_B$, yielding $R^*$=6017 $a_0$. 
One can notice how the simple analytic formulas summarized above provide a remarkable description of the low-energy physics obtained from (the much more complex!) CC calculations of the real Li-Cr interaction potential. Similar agreement can be found near isolated FRs of any other two-body system, which makes the zero-range approximation a valuable simple tool, able to quantitatively reproduce the physics of resonantly interacting  atomic gases at ultralow temperatures.      
 
\subsection{Three-fermion systems within the Born-Oppenheimer approximation} 
\label{3B}
Let us now move to discuss how the introduction of an additional (identical) heavy particle $M$ modifies the two-body scenario above discussed. 
Following Refs.~\cite{Levinsen2011,Petrov2012}, for $M/m\gg 1$, the physics of three-body systems can be qualitatively understood within the Born-Oppenheimer (BO) approximation \cite{B-O_App1927,Fonseca1979}. By taking advantage of the large mass asymmetry, we can assume that the state of the light particle -- at position $\textbf{r}$ -- adiabatically adjusts itself to the relative distance $\textbf{R}$ between the two heavy atoms. 
By further assuming that the heavy-light interaction is entirely characterized by the $s$-wave scattering amplitude Eq.~(\ref{f0}), and that $a\!>0$, the ground state of the three-body system at asymptotically large distances ($R\!\rightarrow\!\infty$) can be considered as being composed by a heavy particle plus a heavy-light dimer. 
As in the double-well problem with tunneling, the state localized near one heavy atom is mixed with the state localized near the other and, correspondingly, the wavefunction describing the motion of the light particle can be written as (see Eq.~(\ref{pole}))
\begin{equation}
 \label{psiBO}
\psi_{\textbf{R},\pm}(\textbf{r}) \propto \frac{e^{-\kappa_{\pm}(R)|\textbf{r}-\textbf{R}/2|}}{|\textbf{r}-\textbf{R}/2|} \pm \frac{e^{-\kappa_{\pm}(R)|\textbf{r}+\textbf{R}/2|}}{|\textbf{r}+\textbf{R}/2|}.
\end{equation}
The corresponding energy,
\begin{equation}
 \label{epsilonBO}
\epsilon_{\pm}(R) = -\frac{\hbar^2 \kappa_{\pm}^2(R)}{2 m},
\end{equation}
is obtained by imposing the Bethe-Peierls boundary conditions Eq.~(\ref{BP}) at $|\textbf{r} \pm \textbf{R}/2|\!\rightarrow$0 (see Ref.~\cite{Levinsen2011} for details) which, similarly to Eq.~(\ref{pole}), yields an equation for $\kappa_{\pm}(R)$,
\begin{equation}
 \label{kappaBO}
\frac{1}{a}-R^* \kappa_{\pm}^2(R)-\kappa_{\pm}(R)=\pm \frac{e^{-\kappa_{\pm}^2(R) R}}{R},
\end{equation} 
that can be solved numerically, thus providing $\epsilon_{\pm}(R)$ through Eq.~(\ref{epsilonBO}). 
The latter then acts as an effective potential within the Schr{\"o}dinger equation for the wavefunction $\phi(\textbf{R})$ describing the motion of the heavy particles.

\begin{figure}
\centering
\includegraphics[width=.8\columnwidth]{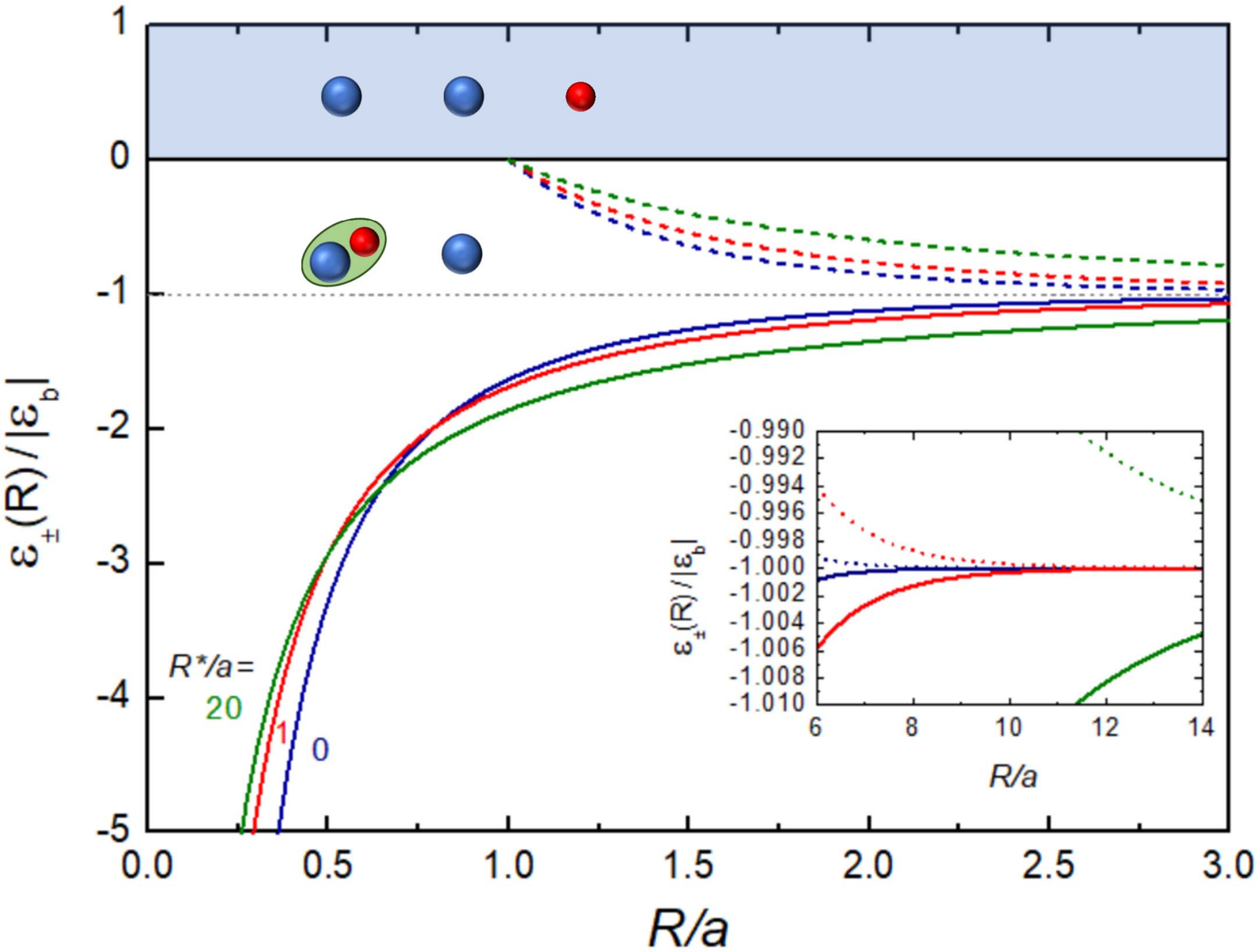} 
\vspace*{-10pt}
\caption{Effective interaction potentials $\epsilon_{\pm}(R)$ are plotted as a function of $R/a$ for different $R^*/a$ values, see legend. 
$\epsilon_{\pm}(R)$ are normalized to the dimer energy which, in the BO approximation, where $m_r\!=m$, reads $\epsilon_b\!=-\hbar^2/(2 m a^{*2})\!=\epsilon_{\pm}(\infty)$. While $\epsilon_{+}(R)$ corresponds to a purely attractive potential, $\epsilon_{-}(R)$ is purely repulsive, and it reaches the three-atom, zero-energy threshold for $R\!=\!a$. For $R\!\gg\!a$, see inset, all curves tend to the bare dimer energy, which sets the atom-dimer scattering threshold. A similar figure and a more detailed discussion is found in Ref.~\cite{Petrov2012}.}  
\label{epsilon_pm}
\vspace*{-20pt}
\end{figure}
Examples of $\epsilon_{+}(R)$ (solid lines) and $\epsilon_{-}(R)$ (dashed lines), normalized to the corresponding dimer binding energy $|\epsilon_b|$, are shown in Fig.~\ref{epsilon_pm} as a function of $R/a$ for different $R^*/a$ values, see legend. 
One can notice the following important features: 
First, although originating from a short-range heavy-light interaction, the effective potentials $\epsilon_{\pm}(R)$ are \textit{long-ranged}: only for $R\!\gg\!a$ (see also inset), all curves tend to $-1$, i.e. to the bare binding energy of one dimer $\epsilon_{\pm}(\infty)\!=\!\epsilon_b$, which in turn corresponds to the atom-dimer scattering threshold. 
Second, while $\epsilon_+(R)<\epsilon_{\pm}(\infty)$ at all distances, thus yielding a purely \textit{attractive} potential, $\epsilon_-(R)$ corresponds to a net effective \textit{repulsion} between the heavy particles. 
In particular, $\epsilon_-(R)$ reaches the three-atom continuum (zero energy) at $R\!=\!a$, irrespective of the $R^*/a$ value (you can see this by considering that $\kappa_-(R)=0$ solves Eq.~(\ref{kappaBO}) for $R=a$). Third, for $R/a\ll 1$ and small $R^*$ values, for which the right-hand side of Eq.~(\ref{kappaBO}) can be neglected, $\epsilon_+(R)$ is determined by $\kappa_+(R) R\!=\!\alpha$=0.56714..., solution of the equation    
$\alpha\!=\!e^{-\alpha}$. Correspondingly,
\begin{equation}
 \label{epsilon_plus}
\epsilon_+(R) \sim -\frac {0.161 \hbar^2}{m R^2} 
                     \tx{ for\ } R\ll a.
\end{equation} 
Depending whether the heavy particles are identical bosons or fermions, the three-body wavefunction, proportional to the product $\psi_{\textbf{R},\pm}(\textbf{r}) \phi(\textbf{R})$, must be respectively symmetric or anti-symmetric with respect to the permutation $\textbf{R} \leftrightarrow -\textbf{R}$. 
Since $\psi_{\textbf{R},+}(\textbf{r})$ is symmetric ($\psi_{\textbf{R},-}(\textbf{r})$ is anti-symmetric) with respect to particle permutation, for the case of heavy fermionic atoms we are interested in here, $\epsilon_+(R)$ ($\epsilon_-(R)$) will act on odd (even) partial-wave channels: 
Solely due to the composite nature of the dimer, combined with quantum statistics of the heavy particles, the atom-dimer interaction thus becomes channel-dependent, the symmetric (anti-symmetric) state corresponding to odd (even) values of the total angular momentum \textit{l}. In spite the two-body interaction is only determined by the scattering amplitude $f_0(k)$ of Eq.~(\ref{f0}),  several partial waves in Eq.~(\ref{ftot}) are required to properly describe the corresponding atom-dimer scattering.

In particular, the $s$-wave channel for the atom-dimer collisions is characterized by a net repulsive potential $U_{0}(R)\!\equiv\!\epsilon_{-}(R)-\epsilon_{\pm}(\infty)$. This features an overall barrier height equal to the bare dimer energy $\epsilon_{\pm}(\infty)$, which in turn depends upon the $R^*/a$ value, see Fig.~\ref{epsilon_pm} and Eq.~(\ref{Eb}). 
The $U_{0}(R)$ repulsion causes the fermion-dimer scattering length $a_{FD}$ to be always positive, and on the order of the atom-atom one, see Ref.~\cite{Levinsen2011,Petrov2003}. 
In this respect, it should be remarked the stark difference with the two-body problem discussed in the previous section: There, for short-range interatomic potentials, an $a\!>$0 value inherently requires the presence of a  bound (dimer) state  below the scattering threshold, see Eq.~(\ref{f0vsfBW}) and related discussion. Here, instead, where the effective potential $U_0(R)$ is long-ranged, the scattering length $a_{FD}$ is positive, although no $s$-wave bound (trimer) state  exists. 
Note that this behavior qualitatively holds also for the case of scattering of dimers with light fermions, although this case cannot be legitimately described through the BO approximation.
\begin{figure}
\centering
\includegraphics[width=.8\columnwidth]{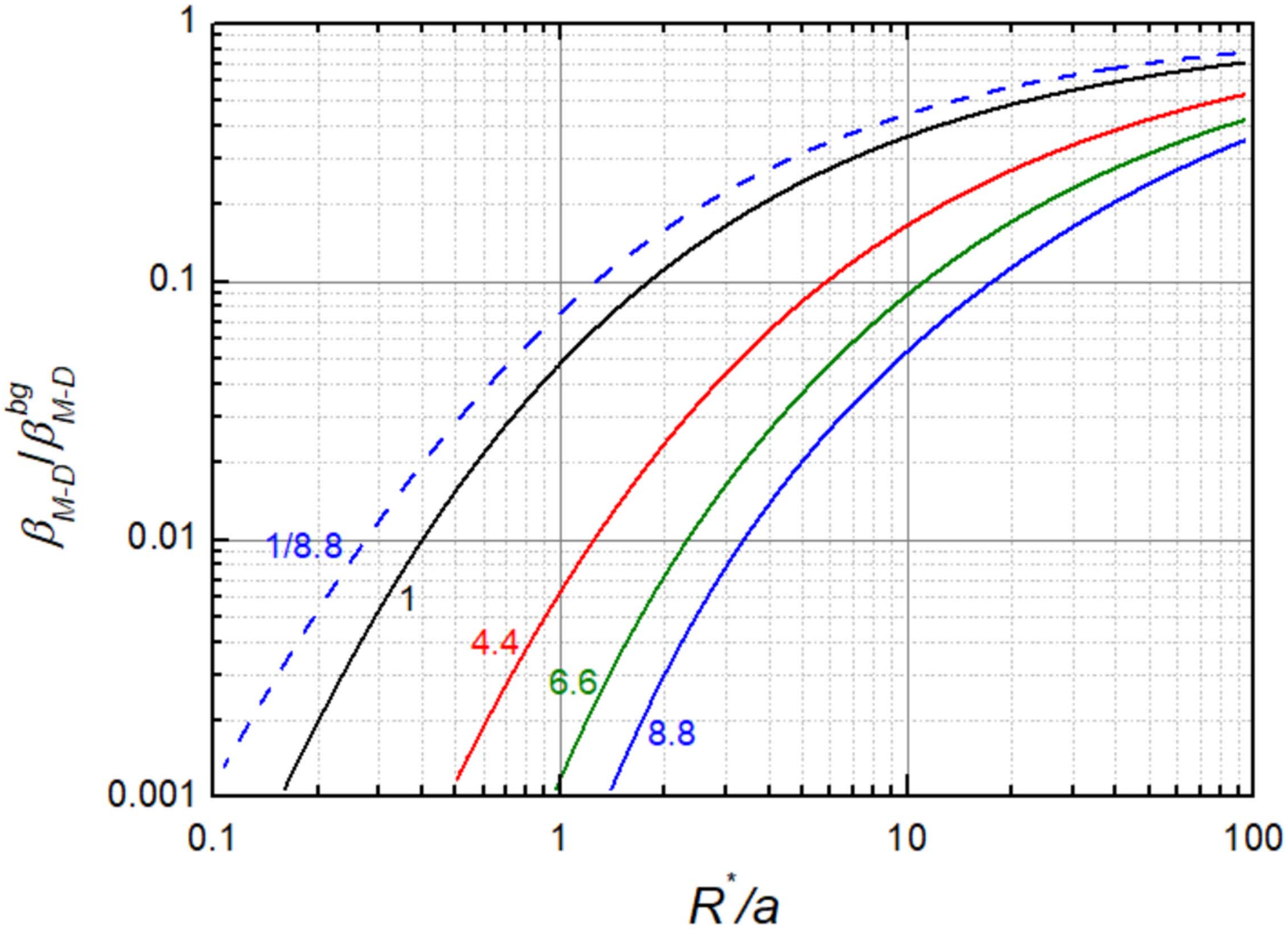} 
\vspace*{-10pt}
\caption{Suppression factors $\beta_{M-D}/\beta_{M-D}^{bg}$ for inelastic atom-dimer scattering, evaluated as a function of $R^*/a$ for different mass ratios, see legend.
Data are taken from Ref.~\cite{Jag2016} and are here combined within one single plot for comparison. The chosen mass ratios are relevant for atomic mixtures of experimental interest: equal-mass systems, Dy-K, K-Li and Cr-Li. The dashed line provides an example of the suppression factor for dimer scattering with light atoms, and the chosen mass ratio of $1/8.8$ is relevant for collisions between Li atoms and LiCr dimers. For increased mass asymmetry, $\beta_{M-D}/\beta_{M-D}^{bg}$ is substantially reduced in the heavy-heavy-light case, while it is only moderately enhanced in heavy-light-light systems, relative to $M/m=$1.     
}  
\label{etaAD}
\vspace*{-15pt}
\end{figure}

Additionally, the presence of a repulsive barrier in $\epsilon_{-}(R)$ at $R\!\geq\!a$, through which low-energy particles have to tunnel to approach each others at short distances $R\sim R_e$, qualitatively explains the celebrated collisional stability of strongly-interacting fermionic systems against inelastic processes \cite{Levinsen2011,Petrov2012, Petrov2005}. 
Specifically, since $U_0(R)\! \propto 1/m$, whereas the kinetic energy of the heavy particles scales as $1/M$, the larger the mass ratio, the more the rate constant $\beta_{M-D}$ for inelastic $M\!-D$ atom-dimer processes in the $s$-wave channel is suppressed, with respect to its background, off-resonant value $\beta_{M-D}^{bg}$ (see Ref.~\cite{Jag2016} for a detailed discussion). 
On the other hand, an increased effective range value effectively weakens such a potential barrier,  owing to the fact that the dimer energy, at given $a$, is reduced by an increased $R^*/a$ value, according to $\epsilon_b(R^*/a\! \gg 1)\!/\epsilon_b(R^*\!=0)\!\sim (a/R^*)$, see Eq.~(\ref{Eb}). For this reason, the stability of fermionic systems near narrow FRs is reduced, with respect to the broad resonance case. 
This trend is illustrated in Fig.~\ref{etaAD}, which combines exact calculations, given in Ref.~\cite{Jag2016}, of $\beta_{M-D}\!/\beta_{M-D}^{bg}$ as a function of $R^*/a$ for various mass ratios, see legend.
One can notice how a given suppression factor is reached for significantly larger $R^*/a$ values as $M/m$ is  increased: E.g. $\beta_{M-D}/\beta_{M-D}^{bg}\!\leq$0.1 requires $R^*/a\!\leq$1 for equal-mass systems, whereas this is obtained already  at $R^*/a\sim$1 for $M/m\!>$6. On the other hand, for inelastic scattering of dimers with light atoms -- i.e. $M/m<$1, see dashed blue line -- the suppression factor is only weakly reduced, relative to the equal-mass case. Although not discussed here in any detail, it is important to emphasize that this picture qualitatively holds also when dimer-dimer collisions are considered, both for what concerns elastic and inelastic $s$-wave scattering.

\begin{figure}
\centering
\includegraphics[width=.8\columnwidth]{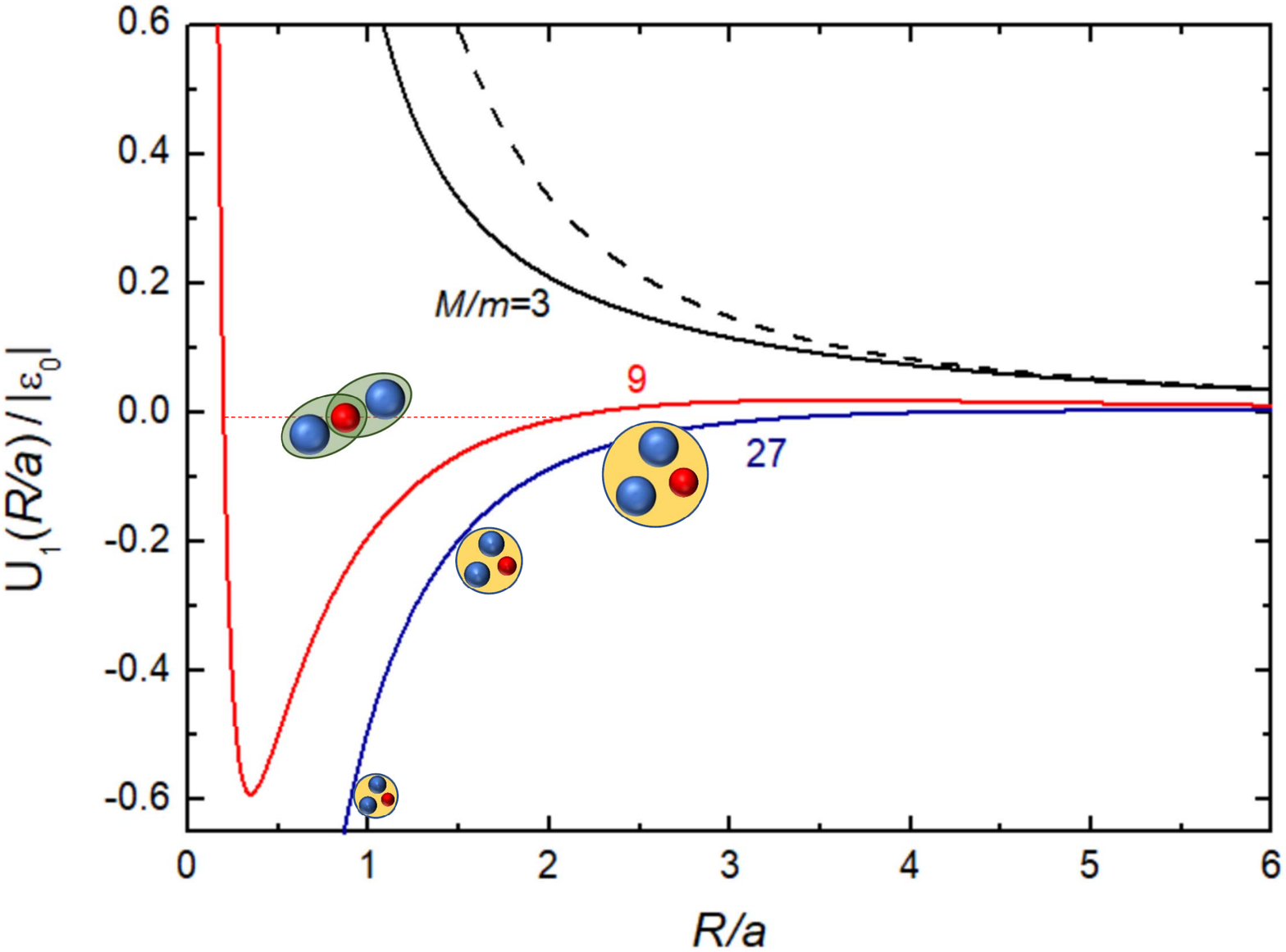} 
\vspace*{-10pt}
\caption{Total effective potentials $U_1(R)=(\epsilon_+(R)-\epsilon_+(\infty))+U_{cb}(R)$, evaluated for $R^*\!=0$, are plotted as a function of $R/a$ for different mass ratios, see legend. The curves are normalized to the zero-range dimer energy $|\epsilon_0|$.
For small $M/m$ values, see black line, $U_1(R)$ is only slightly weaker than the bare centrifugal potential $U_{cb}(R)$, see dashed line. By contrast, when the mass ratio is increased, $U_1(R)$ first develops a potential well at large distances $R\!\sim\!a$, which can support KM trimer states for 8.17$\leq\!M/m\!\leq$13.6. For mass ratios exceeding 13.6, $U_{cb}(R)$ is overcome by the induced attraction $\epsilon_+(R)-\epsilon_+(\infty)$ at all distances, resulting in a purely attractive $1/R^2$ potential, responsible for the appearance of the Efimov effect.}  
\label{BOpotMsum}
\vspace*{-15pt}
\end{figure}

In the odd channels, and primarily in the $p$-wave one, the attractive $\epsilon_+(R)$ competes with the centrifugal barrier $U_{cb}(R)= + \hbar^2 l (l+1)/(M R^2)$. 
Let us first focus on the $R^*=0$ case: Given that from Eq.~(\ref{epsilon_plus}) $\epsilon_+(R) \propto - 1/(m R^2)$, whereas $U_{cb} \propto + 1/(M R^2)$, it is clear that qualitatively different scenarios may arise depending on the mass ratio $M/m$ among the two system components. 
Figure~\ref{BOpotMsum} shows the total effective potential $U_{1}(R) \equiv (\epsilon_+(R)-\epsilon_+(\infty))+U_{cb}(R)$, normalized to the two-body binding energy $|\epsilon_{\pm}(\infty)|$, for different mass ratios, see legend. 
For large mass asymmetries, $M/m>$13.6, the induced interaction overcomes the centrifugal barrier at all distances (see blue line), leading to an overall $1/R^2$ attractive potential independent on $a$, see Eq.~(\ref{epsilon_plus}). 
This is responsible for the appearance of the Efimov effect \cite{Efimov1970,Efimov1973}, with $U_{1}(R)$ supporting an infinite ladder of trimer states as $|a|\!\rightarrow\!\infty$. 
At intermediate mass ratios,  8.17$<M/m<$13.6, $U_{1}(R)$ develops a potential well at large distances $R\! \sim a$ (see red line), which supports at most two \textit{non-Efimovian} trimer states. The latter, not experimentally revealed in any physical system thus far, have been theoretically discovered by Kartavtsev and Malykh about fifteen years ago \cite{Kartavtsev2007} -- although indirect signatures of such exotic states could be inferred from a non-trivial dependence upon $M/m$ of the three-fermion recombination rate at $a>0$, previously identified in Ref.~\cite{Petrov2003}.
As $M/m\!<$8.17, the potential well becomes too shallow to support any bound state, and the last Kartavtsev-Malykh (KM) trimer turns into a ($p$-wave) scattering resonance into the atom-dimer continuum \cite{Kartavtsev2007,Endo2012,Levinsen2011} -- a phenomenon that we first revealed in  Innsbruck  on $^6$Li-$^{40}$K mixtures  \cite{Jag2014}, for which $M/m\!\sim$6.64: 
As in the Feshbach resonance case, the presence of a virtual (trimer) state closely above the collision threshold leads to a strong atom-dimer \textit{attraction}. 
This can overcome the competing repulsive interaction in the $s$-wave channel, causing both a collision-induced \textit{red-shift} and a broadening of the $^{40}$K atomic lines, see Eqs.~(\ref{scatrate}) and (\ref{enshift}), that we could spectroscopically determine \cite{Jag2014}.
This atom-dimer resonance rapidly widens as the mass ratio is further reduced until, for small $M/m$ values (see black solid line), the induced attraction only weakly perturbs the centrifugal barrier (black dashed line). 

It is important to emphasize the qualitatively different nature between Efimov states and KM trimers, which can be appreciated by simply looking at the corresponding BO potentials: The former ones originate by the $1/R^2$ attraction, which extends over all length-scales for $|a|\!\rightarrow \infty$, irrespective of the sign of $a$. Furthermore, as nothing prevents the three-atom wavefunction to extend down to arbitrarily short distances,  $R\!\sim R_e\!\ll a$, Efimovian trimers are inherently unstable against inelastic relaxation (in the $p$-wave channel) towards deeply-bound molecular states. 
By contrast, the KM trimers exist only for $a>0$ values, as a result of the peculiar competition between the centrifugal barrier and the induced attraction $(\epsilon_+(R)-\epsilon_\pm(\infty))$. Similarly to Feshbach dimers, KM trimers are universal halo states, whose energy and size, for given mass ratio, are solely determined by the scattering length $a$. 
Most importantly, they are expected to be collisionally stable, thanks to the presence of the potential barrier that prevents particles to approach each other at short distances, see red line in Fig.~\ref{BOpotMsum}. This makes the KM trimers, as well as the associated atom-dimer $p$-wave resonances, extremely appealing also from a many-body perspective, as they represent a unique example -- within the field of ultracold gases -- of a non-perturbative few-body phenomenon with purely elastic character.

\begin{figure}
\centering
\includegraphics[width=.8\columnwidth]{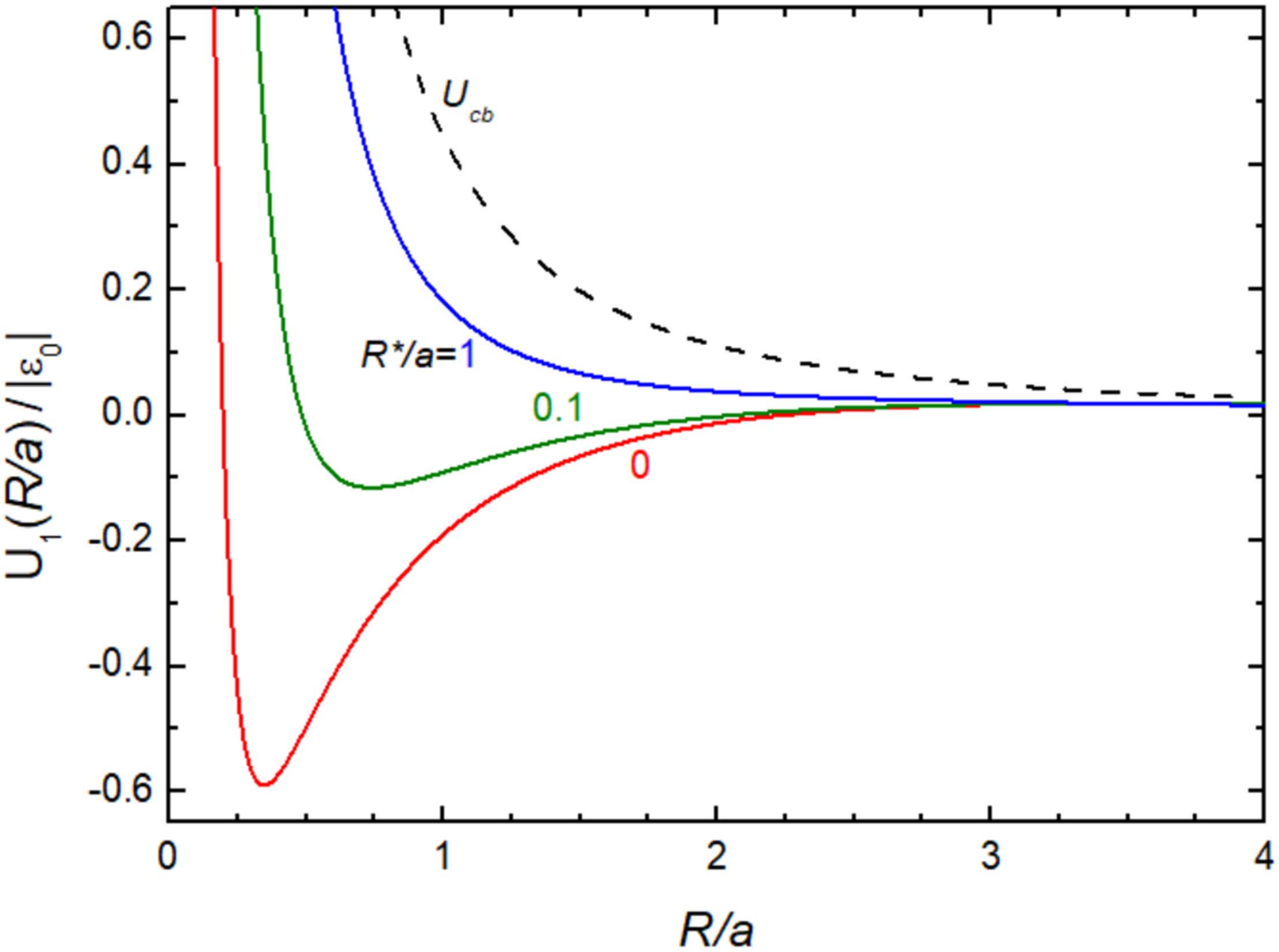} 
\vspace*{-10pt}
\caption{Total effective potentials $U_1(R)=(\epsilon_+(R)-\epsilon_+(\infty))+U_{cb}(R)$, evaluated for $M/m=9$, are plotted as a function of $R/a$ for different $R^*/a$ values, see legend. As for Fig.~\ref{BOpotMsum}, the curves are normalized to the zero-range dimer energy $|\epsilon_0|$.
Since $\epsilon_-(R)$ changes from a $1/R^2$ attractive potential at $R^*/a\!=$0, to a $1/R$ Coulomb-like trend for $R^*/a\!\gg$1, the contribution of the centrifugal barrier (dashed line) becomes progressively stronger with increasing $R^*/a$ values, sensitively weakening $U_1(R)$ with respect to the zero-range limit.
}  
\label{BOpotRstar}
\vspace*{-15pt}
\end{figure}
Let us  finally consider how a finite effective range parameter affects the scenario above discussed for the $p$-wave channel. To this end, Fig.~\ref{BOpotRstar} shows the total effective potential $U_1(R)$, evaluated for $M/m$=9 and different $R^*/a$ values, see legend.
One can notice how already a small but finite effective range sensitively modifies the shape of $U_1(R)$, the depth of the potential well being strongly reduced for increasing $R^*/a$ values. 
Such a feature can be understood by considering again the trends of $\epsilon_-(R)$, obtained from Eq.~(\ref{kappaBO}) and presented in Fig.~\ref{epsilon_pm}: For $R^*$=0, we already saw that $\epsilon_-(R\ll a)$ is characterized by the peculiar $R^{-2}$ scaling -- identical but opposite to that of the centrifugal barrier, see  Eq.~(\ref{epsilon_plus}) and related discussion. In the opposite limit $R^*/a \gg 1$, instead, one can easily verify that the induced attraction behaves like 
\begin{equation}
 \label{epsilon_plusRstar}
\epsilon_+(R)|_{R^*/a\!\gg 1} \sim -\frac {\hbar^2}{2 m R^* R} 
                     \tx{ for\ } R\ll R^*.
\end{equation}      
Namely, for large $R^*/a$ values, $\epsilon_+(R)$ is a \textit{Coulomb-like} potential, and thus the Schr{\"o}dinger equation for the heavy fermions is governed by a hydrogen-like Hamiltonian, characterized by an `effective Bohr radius' $\propto R^*/(M/m)$. 
This makes that, for increasing $R^*$ values, the effect of the $p$-wave centrifugal barrier  becomes progressively more relevant within $U_1(R)$, an increased value of $R^*/a$ playing qualitatively the same role of a decreased mass ratio $M/m$, see Figs.~\ref{BOpotMsum} and \ref{BOpotRstar} comparison.  
Such a sensitive $R^*/a$ dependence was indeed observed in the $^6$Li-$^{40}$K Innsbruck experiment \cite{Jag2014} -- see also R. Grimm's contribution to these Proceedings -- for which $R^*\sim$2700 $a_0$ \cite{Kohstall2012, Cetina2015}: For small magnetic-field detuning, where $R^*/a\leq$1, a strong $p$-wave attraction dominated  the K-LiK scattering, yielding an overall red shift $\delta \nu\!<$0 of the spectroscopy lines, see Eq.~(\ref{enshift}). By moving away from the K-Li FR -- i.e. progressively increasing $R^*/a$ -- this effect was first canceled, and then overcome, by the repulsion in the $s$-wave channel.
\vspace{15pt}

In conclusion of this part, it is useful to recap the main results discussed herein:
\begin{itemize}
\item Although two-body interactions between ultracold atoms are short-ranged in nature, the corresponding three-body system exhibits an inherently long-range, $l$-dependent interaction, which requires several partial-wave contributions to Eq.~(\ref{ftot}) to be properly described. 
\item The peculiar trends of $\epsilon_{\pm}(R)$ obtained within the BO approximation qualitatively explain some well-known (but often mysterious) results of few-body physics.
In particular,  for fermionic particles, the $s$-wave channel is characterized by a repulsive $\epsilon_-(R)$ potential, yielding a positive atom-dimer elastic scattering length, and causing a suppression of inelastic scattering. 
\item In the $p$-wave channel, the induced attraction $\epsilon_+(R)$ concurs with the centrifugal barrier, resulting in the total effective potential $U_1(R)$. Owing to the different dependence of the two potential terms upon the masses of the two system components, the resulting three-body properties sensitively depend on the mass ratio of the system: In particular, Efimov (KM universal) trimers can exist for large $M/m>$13.6 (intermediate 8.17$<M/m<$13.6) values in the $R^*$=0 limit. 
\item Efimovian and KM scenarios  starkly differ also for what concerns inelastic scattering: In the Efimov case, $U_1(R)$ is attractive at all distances, favoring an enhancement of inelastic processes. In contrast, for $M/m<$13.6, the system is protected against inelastic collisions by the centrifugal barrier dominating at short distances.  
\item Finite-range effects sensitively modify the three-body system properties. Specifically, an increased $R^*/a$ value reduces the stability against  $s$-wave inelastic collisions, and it weakens the attraction in the $p$-wave channel, an increased value of the $R^*/a$ parameter playing qualitatively the same role of a decreased mass ratio $M/m$ in the zero-range limit.               
\end{itemize}

Finally, a few additional comments are also due for what concerns (i) the role of a reduced dimensionality, and (ii) the effect of adding more (identical heavy) particles to the few-body system.
Not entering into any detail while referring the reader to Refs.~\cite{Blume2012,Nishida2008,Levinsen2009,Ngampruetikorn2013,Levinsen2013,Bazak2017b, Liu2022A}, it is important to remark that confining the particles into a reduced geometry generally favors their clustering. This trend can be understood by considering the peculiar competition between $\epsilon_-(R)$ and $U_{cb}(R)$, discussed above for the three-dimensional (3D) case: Since, e.g., passing from three to two dimensions the centrifugal barrier is reduced from $U_{cb}^{3D}(R)\!=\!\hbar^2 l (l+1)/(M R^2)$ to $U_{cb}^{2D}(R)\!=\!\hbar^2 l^2/(M R^2)$ -- i.e. by a factor of 2 for the $p$-wave channel-- a smaller mass ratio of $M/m\!\sim$3 is there sufficient for the emergence of a bound-state trimer, see Ref.~\cite{Pricoupenko2010}.
The emergence of $N>$3-body cluster states, made of one light particle plus $N\!-$1 heavy fermions, is also possible, although it requires an increased $M/m$ value, relative to the three-body case. For instance, 3D universal $p$-wave tetramers (pentamers) are expected to appear, in the zero-range limit, for $M/m\!\geq$8.886 and $M/m\!\geq$9.672, respectively \cite{Bazak2017}.

\section{The special case of lithium-chromium: theoretical considerations} \label{SecLiCrTheo}

In the previous Section we have seen how the few-body physics of resonantly interacting fermions can be considerably enriched, with respect to the equal-mass case, when a sufficiently large mass asymmetry is introduced among the system constituents.
In the following, I highlight the special properties of ultracold lithium-chromium ($^6$Li-$^{53}$Cr) Fermi mixtures, and their potential interest for future few- and many-body studies.
As discussed in Sec.~\ref{3B}, fermionic systems characterized by intermediate mass ratios comprised in the window 8.17$\leq M/m \leq$13.6, are expected to exhibit a very peculiar few-body physics, starkly distinct from the widely-explored Efimov scenario: While elastic $p$-wave three-body interactions can be substantial  for these mixtures, inelastic processes, typical of Efimovian systems, are here absent. This makes such an intermediate regime of mass ratios a potentially unique playground that potentially allows one to access a yet unexplored branch of few-body physics and, even more interestingly, to investigate how non-perturbative few-body effects modify the many-body properties of highly-correlated fermionic matter.

Among the possible combinations of stable fermionic isotopes that -- at least in single-species experiments -- have been already experimentally produced in the ultracold regime, only $^6$Li-$^{53}$Cr ($M/m\sim$8.8) and metastable $^3$He$^*$-$^{40}$K ($M/m\sim$13) mixtures fall in the interesting intermediate region. 
The mass ratio of the $^{40}$K-$^6$Li system, $M/m\sim$6.6, although being too small to support a stable KM trimer state in three dimensions, is in principle also appealing, as it supports -- in the $R^*$=0 limit -- a $p$-wave resonance in K-KLi scattering at rather low collision energies, $E_{coll}\sim$0.1 $|\epsilon_0|$ \cite{Levinsen2011}. Moreover, the scattering resonance could be turned into a real trimer when confining the system in a quasi-2D geometry \cite{Levinsen2009}.
Unfortunately, though, the widest K-Li Feshbach resonances occur in excited hyperfine state combinations that are not immune to  two-body decay processes \cite{Wille2008,Naik2011}, which inherently limit the system lifetime under resonant conditions. Furthermore, all K-Li  FRs are narrow in nature \cite{Naik2011}: As a result, while a sizable K-KLi attraction could be experimentally revealed and characterized \cite{Jag2014}, the large effective range of this mixture ($R^* \sim$2700 $a_0$)  considerably weakens the maximum  strength of the atom-dimer interaction attainable in experiments, compared with the one predicted in the zero-range limit \cite{Levinsen2011, Jag2014}.

\begin{figure}
\centering
\includegraphics[width=.8\columnwidth]{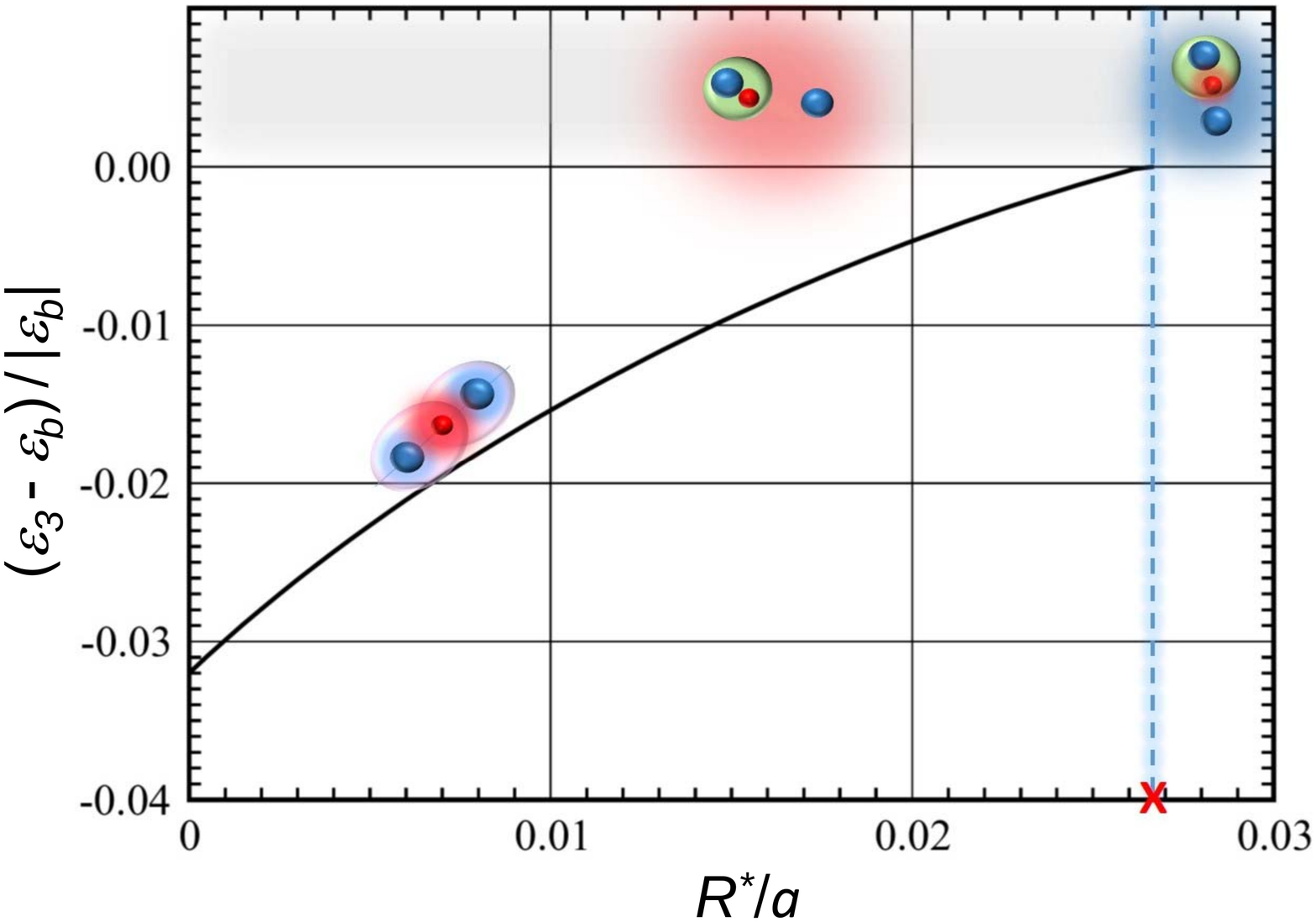} 
\vspace*{-10pt}
\caption{Theory prediction for the LiCr$_2$ trimer binding energy as a function of $R*/a$. $\epsilon_3-\epsilon_b$ represents the energy detuning of the three-body bound state, relative to the atom-dimer energy threshold $\epsilon_b$.  In the broad resonance limit $R^*$= 0, and up to a critical value $(R^*/a)_c\sim$ 0.027 (marked by the red cross), a LiCr$_2$ trimer is stable in three dimensions. Accordingly, below $(R^*/a)_c$, a strong repulsion occurs between Cr atoms and LiCr dimers. Once the trimer energy reaches the atom-dimer threshold, the atom-dimer interaction becomes resonant, and above the critical point a strong $p$-wave attractive interaction between Cr and LiCr develops (D. S. Petrov, Private Communication 2014).
}  
\label{Cr2Li}
\vspace*{-15pt}
\end{figure}

If this fact is annoying for the Li-K mixture, it emerges as a formidable tool for the Cr-Li system, whose mass ratio is sufficiently large for a stable LiCr$_2$ KM trimer to exist. 
The energy of such a state is   only 3\% lower than the dimer energy, hence making the system extremely sensitive to any deviation from the zero-range potential. In particular, already a small yet finite value of $R^*/a\sim$0.03 lifts the trimer state energy $\epsilon_3$ up to the atom-dimer threshold, see Fig.~\ref{Cr2Li}, leading to the occurrence of a $p$-wave scattering resonance between Cr and LiCr dimers. 
Notably, the resonant dependence of the atom-dimer scattering properties on $R^*/a$ directly turns into a dependence on the magnetic-field detuning from the Li-Cr resonance pole. This uniquely enables a resonant tuning of \textit{both} two- and three-body \textit{elastic} interactions in the experiment, irrespective of the specific $R^*$ value associated with lithium-chromium FRs. 

Furthermore, as already mentioned at the end of Sec.~\ref{3B}, also a universal (bosonic) tetramer is predicted to exist in three dimensions for $M/m\geq$8.86 \cite{Bazak2017}: Since the Cr-Li mass ratio is less than 1\% below such critical value, it implies that a LiCr$_3$ tetramer is essentially resonant with the (two-)atom-dimer threshold, and that this could be easily stabilized by  applying a moderate quasi-2D confinement.
The lithium-chromium mixture thus emerges as a unique platform for the experimental investigation of non-Efimovian few-body cluster states and scattering resonances, which could be experimentally revealed based on the same spectroscopic protocols we already developed for the study of Li-K mixtures \cite{Jag2014} -- see R. Grimm's contribution to these Proceedings.

As already mentioned, such non-Efimovian few-body effects are purely elastic in nature, which makes Li-Cr mixtures extremely appealing also for the exploration of new exotic many-body regimes, triggered by non-perturbative few-body effects, and  essentially unexplored even at the theoretical level, yet.
Indeed, how the phase diagram of this system will look like is a completely open question, but it is reasonable to expect that this may host a variety of highly non-trivial normal and superfluid states. 
The wealth of such many-body system can be inferred from the exploration of the `light impurity' problem \cite{Mathy2011, Massignan2014,Liu2022B,Scazza2022}, see also M. Parish's contribution to these Proceedings: Indeed,  this regime already provides important information about the phases of the system at finite polarizations, the impurity limit exhibiting some of the critical points of the full zero-temperature phase diagram \cite{Parish2007}. Experimentally, this scenario can be realized by embedding a few Li atoms within a degenerate chromium Fermi gas, that may be subsequently  probed via radio-frequency spectroscopy protocols \cite{Kohstall2012, Massignan2014,Scazza2022},  as well as with state-of-the-art Bragg and Raman spectroscopy schemes \cite{Veeravalli2008,Ness2020}. 
For instance, in contrast with so-far investigated fermionic systems \cite{Schirotzek2009, Kohstall2012, Koschorreck2012, Scazza2017}, three rather than two kinds of quasi-particles are expected to exist already for $M/m\leq$7: polarons, dressed dimers, and trimers, respectively \cite{Mathy2011}. 
In particular, the dimer may be found to be the ground state, either at zero or finite momentum, the latter case being the impurity-limit analog of a FFLO superfluid phase at lower polarization \cite{Mathy2011}. 
Similarly, a stable trimer quasi-particle points to the possible emergence of a (three-component!) Fermi gas of $p$-wave three-body clusters at finite impurity densities \cite{Endo2016}. 
Even more exotic scenarios can be envisioned, where bosonic tetramers could condense into a superfluid state with highly non-trivial topological properties, or where a host LiCr Bose gas could foster boson-mediated interactions and exotic ($p$-wave) pairing of fermionic Cr atoms \cite{Bulgac2006,Bulgac2009}. 

From this perspective, it is interesting to notice the similarity shared between the Li-Cr atomic mixture -- eventually confined in a quasi-2D geometry -- and electronic matter within 2D transition metal dichalcogenides (TMDs): There, electrons and holes can pair up to form tightly-bound excitons, the solid-state analog of LiCr dimers.  Additionally, these materials also support the existence of trions \cite{Courtade2017} --
 exciton-electron bound states which exhibit binding energies only slightly lower than the exciton one, thus qualitatively playing the role of KM trimers in the solid-state context. 
The presence of trions provides a strong non-linear exciton-electron coupling, which can be properly tuned upon mechanical deformations and elastic strain, similarly to the effect of a magnetic field tuned around a Cr-dimer resonance in the cold-atom setting.
The resemblance of few-body phenomena already observed in TMDs with those predicted in atomic Fermi mixtures, suggests that also similar many-body regimes may develop within these two environments, and a growing cross-fertilization between the two fields has been recently  triggered:
For instance, attractive and repulsive Fermi polarons, initially unveiled in cold-atom platforms \cite{Massignan2014, Schmidt2018,Scazza2022}, have recently been reported for TMD environments and analyzed on the basis of essentially identical theoretical methods \cite{Sidler2017}. Moreover, both TMDs \cite{Cotlet2016} and ultracold Fermi mixtures \cite{Bulgac2006,Bulgac2009} are currently considered in the pursuit for novel superfluid/superconducting pairing mechanisms, ideally featuring high-temperature character and non-trivial topological properties.

Lithium-chromium mixtures also appear as very promising candidates for the experimental investigation of many-body regimes of \textit{repulsive} Fermi gases, and of the possible emergence of (ferro-)magnetic phases \cite{Stoner1933}. 
In contrast with the homonuclear systems explored in this context thus far \cite{Jo2009,Sanner2009,Scazza2017,Valtolina2017,Amico2018,Scazza2020} -- for which ferromagnetic correlations develop at sufficiently large repulsion over timescales only slightly faster than those of recombination processes onto the shallow Feshbach dimer state -- for the Li-Cr mixture one expects the pairing instability to be drastically reduced. 
As  briefly  mentioned already in Sec.~\ref{3B}, indeed, the rate for three-body recombination processes of the kind $\uparrow+\uparrow+\downarrow\rightarrow \uparrow+(\uparrow \downarrow)$ exhibits a highly non-monotonic behavior as a function of the mass ratio $m_{\uparrow}/m_{\downarrow}$. 
Specifically, in the $R^*$=0 limit, such processes are exactly \textit{zeroed} for $m_{\uparrow}/m_{\downarrow}$=0 and 8.612, respectively \cite{Petrov2003}, thanks to a purely quantum interference phenomenon, linked to the existence of the first KM trimer state slightly below the atom-dimer threshold. 
In the degenerate regime and wide resonance case, the three-body system comprised of two lithium and one chromium atoms, with $m_{\uparrow}/m_{\downarrow}$=1$/$8.8 (two chromium and one lithium atoms, with $m_{\uparrow}/m_{\downarrow}$=8.8), features a recombination rate about a factor 50 (1700) smaller than the one of equal-mass systems at the same scattering length and densities, i.e. same $\kappa_F a$ value, making the lifetime of Li-Cr mixtures at strong repulsion orders of magnitude longer than that of the equal-mass system.   
Additionally, the critical repulsion $(\kappa_F a)_c$ for ferromagnetic domains to develop is sizably reduced as the mass asymmetry between the two components is increased \cite{Cui2013a}, owing to a reduced Fermi pressure of the heavy fermions. For Li-Cr mixtures, mean-field calculations predict $(\kappa_F a)_c$ to drop by about a factor 2, relative to the $m_{\uparrow}\!=\!m_{\downarrow}$ case. 
Since the dimer formation rate via three-body recombination grows as $(\kappa_F a)^6$ \cite{Petrov2003}, at the critical repulsion for magnetic domains to emerge, a Li-Cr Fermi mixture is expected to exhibit a lifetime against Cr-Cr-Li (Li-Li-Cr) recombination processes that is $\sim 10^5$ (3000) times longer than the one of equal-mass mixtures. This makes the timescale of inelastic dimer formation to lift from a few Fermi times, up to several trap periods, implying that large magnetic domains may develop in Li-Cr mixtures, `immune' from the pairing instability. 
This shapes the lithium-chromium system as a pristine environment for the experimental study of the textbook  Stoner's model for itinerant ferromagnetism \cite{Stoner1933}.

Finally, it is worth mentioning that the Li-Cr system is appealing also from the viewpoint of realizing ultracold gases of polar, paramagnetic molecules, i.e. dimers featuring both electric and magnetic dipole moments.
In fact, recent studies \cite{Zaremba2022} predict LiCr molecules in their absolute ground state to combine a large electric dipole moment, exceeding 3 Debye, with a $S_{LiCr}\!=\!5/2$ electronic spin -- the latter resulting from  the highly-dipolar chromium element ($S_{Cr}\!=$3), combined with the lithium alkali ($S_{Li}\!=\!1/2$). 
With respect to other non bi-alkali combinations -- such as alkali-lanthanide systems \cite{Ravensbergen2020,Green2020,Schafer2021,Barbe2018},  nowadays considered in this context and exhibiting similar ground-state properties -- the Li-Cr system has the great advantage of behaving essentially like a bi-alkali mixture from the view-point of interspecies scattering properties, see also Sec.~\ref{FeshbachExp} below. 
These indeed depend, as for the alkalis, by solely two hyperfine interaction potentials, the $X ^6\Sigma^+$ \textit{sextet} and $a ^8\Sigma^+$ \textit{octet}, respectively, associated with the total electron spin of the atom pair, given by $S_{Cr}\!\oplus\!S_{Li}$=5/2 and 7/2. 
This strong interaction term -- responsible for appearance of the strongest resonances -- combined with the weaker though non-negligible magnetic dipole interaction, leads to several FRs arranged in \textit{non-chaotic} patterns \cite{Ciamei2022B}, in spite of the dipolar nature and complex level structure of fermionic $^{53}$Cr \cite{Neri2020}, reminiscent of $^{161}$Dy \cite{Lu2012} and $^{167}$Er \cite{Aikawa2014}. 
This implies that -- similarly to the bi-alkali case -- full coupled-channel calculations can unambiguously assign a complete set of quantum numbers to experimentally identified FRs  \cite{Ciamei2022B}, and thus to the associated Feshbach dimers. 
This represents a major advantage to experimentally devise efficient optical schemes to coherently transfer weakly-bound LiCr molecules to the ground states of the $X ^6\Sigma^+$ and $a ^8\Sigma^+$ potentials.

\section{$^6$Li-$^{53}$Cr Fermi mixtures in practice: some challenges and a few surprises} \label{SecLiCrExp}

\begin{figure}
\centering
\includegraphics[width=1.\columnwidth]{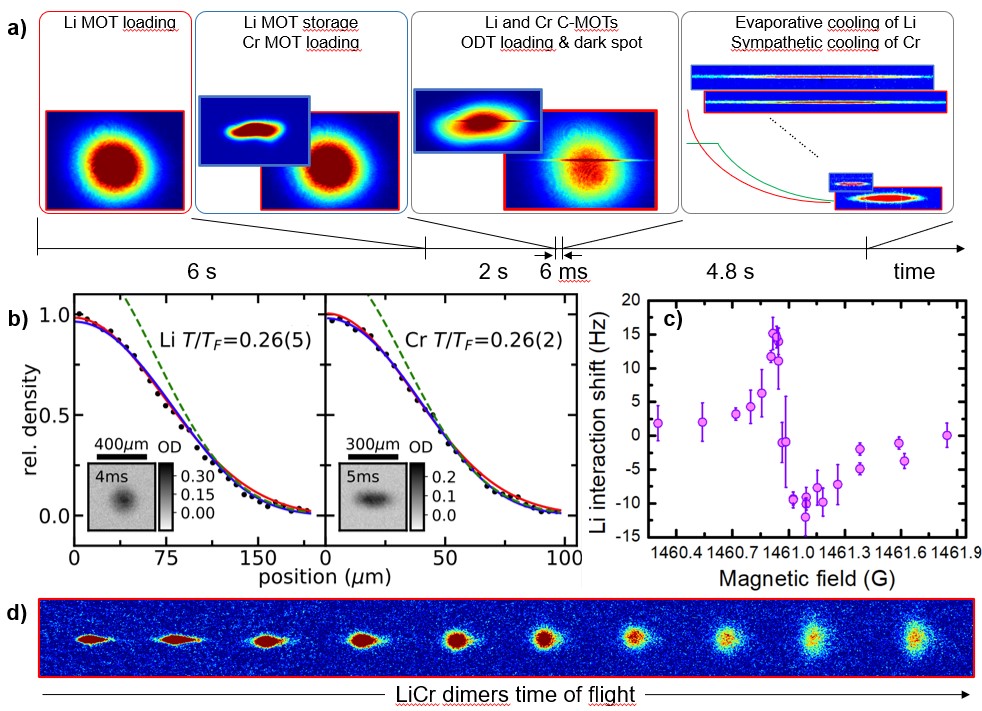}
\vspace*{-15pt}
\caption{
\textbf{a)} Overview of the main steps of our experimental strategy to produce Li-Cr Fermi mixtures, discussed in the text. Typical Li and Cr absorption images, acquired after each stage of the routine are shown, with Li (Cr) images framed in red (blue). The time-line of the experimental cycle is shown below the panels. 
\textbf{b)} Axially-integrated density profiles (black circles) of a lithium (left) and chromium (right) Fermi gas, prepared into the crossed BODT. Profiles are obtained from the average of about 20 independent  images, acquired after time-of-flight expansion, and shown as insets. 
Data are compared with best fits to a Gaussian (red line) and  a Fermi-Dirac (blue line) distribution.
The reduced temperature $T/T_F$ obtained from the latter fit, together with the fit uncertainty, is specified in each panel. A Gaussian fit to the low-density tails of the distributions (green dashed line) accurately captures the atom temperatures while overestimating the central density of the highly-degenerate clouds. 
\textbf{c)} Example of interaction-induced shift of the Li$|1\rangle \!\rightarrow$Li$|2\rangle$ radio-frequency transition, measured with a 100 ms-long spectrocopy pulse on a dilute Li-Cr thermal mixture at 6 $\mu$K, near the Li$|2\rangle$-Cr$|1\rangle$ FR centered at 1461 G \cite{Ciamei2022B}. \textbf{d)} Absorption images, acquired at different time of flights, of a resonantly-interacting cloud of about 5$\times$10$^4$ LiCr dimers at $T<$200 nK, produced through a magneto-association ramp across the 1461 G resonance.   
}
\label{ExpSketch}
\vspace*{-15pt}
\end{figure}

In  Section~\ref{SecLiCrTheo} I have highlighted the peculiar properties theoretically expected for lithium-chromium ($^6$Li-$^{53}$Cr) Fermi mixtures. In the following, I provide a brief overview of the experimental activities concerning the production and characterization of such ultracold system, uniquely available in our lab worldwide, so far. 
In particular, I will discuss how, overcoming a few non-trivial technical challenges -- mainly connected with the limited knowledge of  $^{53}$Cr  in the ultracold regime -- we could devise an all-optical strategy, illustrated in Fig.~\ref{ExpSketch}(a), that allowed us to realize $^6$Li-$^{53}$Cr Fermi mixtures of more than 2$\times 10^5$ $^6$Li atoms and $10^5$ $^{53}$Cr atoms, with both species exhibiting normalized temperatures of about $T/T_{F}$=0.25, see Fig.~\ref{ExpSketch}(b).
Due to length constraints, in Section~\ref{ExperimentalProcedures} I will focus only on those aspects that may be of more general interest, while I refer the interested reader to Refs.~\cite{Simonelli2019,Neri2020,Ciamei2022B,Ciamei2022} for further details on our experimental setup and procedures.
In Section~\ref{FeshbachExp}, I summarize the scattering properties of Li-Cr mixtures detailed in Ref.~\cite{Ciamei2022B}, and highlight a few (yet unpublished) results from ongoing studies of our system near one $s$-wave Feshbach resonance, see Fig.~\ref{ExpSketch}(c)-(d).

\subsection{\textbf{Production of degenerate $^6$Li-$^{53}$Cr Fermi mixtures}}\label{ExperimentalProcedures}
Our strategy to produce degenerate lithium-chromium Fermi mixtures, illustrated in Fig.~\ref{ExpSketch}(a), is conceptually similar to the all-optical one developed for the Li-K system in the Innsbruck experiment \cite{Spiegelhalder2010}, and it consists of the following main steps: (i) Realization of a cold mixture in a dual-species magneto-optical trap (MOT); (ii) Direct loading of the two components into an optical dipole trap (ODT); (iii) Evaporative cooling of a two-state Li mixture, and simultaneous sympathetic cooling of chromium.  
In spite of its conceptual simplicity, successful application of this approach to Li-Cr mixtures has required to tackle various challenges -- mostly connected with the rather-limited experimental investigation of fermionic chromium prior to our studies \cite{Chicireanu2006, Naylor2015}. 

Specifically, a few major issues make the production of ultracold $^{53}$Cr gases non-trivial. 
First, chromium  exhibits a rather complex level structure which requires, besides the main cooling light, up to six repumpers (and three distinct laser sources) to fully close the cooling cycle \cite{Neri2020}. 
At a more fundamental level, chromium suffers from rather strong light-assisted inelastic collisions \cite{Chicireanu2006}, which limited the $^{53}$Cr number collected in the MOT to roughly 10$^5$, before our first attempts \cite{Neri2020}.
Second, direct loading of chromium atoms from the MOT into an infrared (IR) optical dipole trap was proved to be challenging, owing to detrimental light-shifts of the IR light onto the cooling transition \cite{Chicireanu2007,Beaufils2008}.
Finally, efficient sympathetic cooling of chromium with lithium could not \textit{a priori} be taken for granted: 
Although the Li-Cr background scattering length -- initially unknown, and that only recently was found to be about 42 $a_0$ \cite{Ciamei2022B} -- is close to the Li-K one \cite{Wille2008, Naik2011}, and thus sufficient to guarantee a good thermalization rate, efficient Li-Cr sympathetic cooling in a standard IR trap is hard to achieve, given that the chromium polarizability, relative to the lithium one, is about 1$\div$3, in contrast with a potassium-to-lithium polarizability ratio of about 2, around 1 $\mu$m wavelength \cite{Spiegelhalder2010}.
In the following I summarize how these three major issues could be successfully overcome, or circumvented, in our lab. More details are given in Ref.~\cite{Neri2020,Ciamei2022}.

\subsubsection{\textit{Realizing a large Li-Cr double-species MOT}} \label{Li-Cr_MOT}
As the double-MOT is concerned, we first produce a lithium MOT of about $10^9$ atoms within a typical 6 seconds loading time, by following protocols identical to those we already developed for the single-species experiment at LENS, see Ref.~\cite{Burchianti2014}.
Once the lithium component is collected, the MOT parameters are adjusted to their optimal values for Cr loading, see Fig.~\ref{ExpSketch}(a). In contrast with previous studies \cite{Chicireanu2006}, including our first work  Ref.~\cite{Neri2020}, our present strategy -- smartly devised by my colleague Dr. A. Ciamei in spring 2020 -- is based on operating the Cr MOT at very low (normalized) intensity values $s_0=I/I_{sat}\ll$1.     
This, combined with a high atomic flux but with very small exit velocity, of about 15 m$/$s \cite{Neri2020}, delivered from our Cr Zeeman slower, allows us to capture up to 80 millions of chromium atoms in the MOT, while drastically suppressing the rate of light-assisted losses \cite{Ciamei2022}.
The advantage of operating the MOT in this regime can be understood by considering some textbook results  \cite{metcalf1999} about the MOT loading dynamics, generally governed by the rate equation 
\begin{equation}
\frac{dN}{dt}=\Gamma_L -\alpha N(t) - \frac{\beta}{\left\langle V \right\rangle} N^2(t).	
\label{rateeq}
\end{equation}
Here $\Gamma_L$ is  the loading rate, $\alpha$ is a one-body decay rate, $\beta$ is the rate coefficient per unit volume for light-assisted collisions, and $\left\langle V \right\rangle$ denotes the density-weighted volume of the MOT cloud. 
Since we exploit all six repumping lights required for $^{53}$Cr to fully close the cooling cycle, we can safely set $\alpha$=0.
Eq.~(\ref{rateeq}) then yields the asymptotic value for the collected atom number
\begin{equation}
N_{\infty}=\sqrt{\frac{\Gamma_L \left\langle V \right\rangle}{\beta}}.	
\label{Ninf}
\end{equation}
From Eq.~(\ref{Ninf}) one can immediately see that, in order to increase $N_{\infty}$, one needs to maximize $\Gamma_L$ (that depends on the transverse cooling and Zeeman slowing parameters, but not on the MOT ones), enlarge $\left\langle V \right\rangle$ and minimize $\beta$.
Again relying on textbook calculations  \cite{metcalf1999}, one can express $N_{\infty}$ in terms of the MOT parameters -- specifically quadrupole gradient $b$, normalized detuning $\delta/\Gamma$, and normalized intensity $s_0$.
In the limit of $s_0\!\ll$1 and $|\delta|\gg \Gamma$ it is easy to verify that \cite{Ciamei2022}
\begin{equation}
N_{\infty}\propto \frac{\sqrt{\Gamma_L}}{b^{3/4}}\frac{(\delta/\Gamma)^4}{s_0^{5/4}},	
\label{Ninf_asym}
\end{equation}
implying that, for a given loading rate $\Gamma_L$, light-assisted losses can be mitigated -- thereby substantially increasing $N_{\infty}$ -- by working at low $s_0$ values, large detunings, and weak field gradients of the MOT.

A well-known system, where very strong light-assisted losses have been successfully circumvented by following these concepts, is metastable $^4$He$^*$ \cite{Pereira2001,Chang2014}: In that case, the MOT is operated at large detunings, $|\delta|\sim 40\Gamma$, while large $s_0\!>$10 values are maintained to ensure a sufficiently high capture velocity.
In our experiment on the $^{53}$Cr element, which exhibits a saturation intensity (linewidth) more than 50 times (3 times) larger than the one of He$^*$, we cannot reach $s_0\gg$1 without diminishing the performance of transverse cooling and hyperfine pumping stages at the chromium oven \cite{Neri2020}, thus decreasing $\Gamma_L$, owing to the limited amount of blue power available. 
Hence, we followed a strategy opposite to the one of $^4$He$^*$, based on minimizing $s_0$, while keeping relatively small light detunings of a few $\Gamma$.
Experimentally, we could indeed  identify a region of MOT parameters able to strongly mitigate light-assisted losses, allowing us to greatly speed up and simplify the experimental routine to produce a large $^6$Li-$^{53}$Cr mixture in the cold regime: 
Under optimum working conditions, a loading time of 2 s suffices to reach $N_{\infty}$= 8 $\times 10^7$ for chromium \cite{Ciamei2022}, a value about two-to-three orders of magnitude higher than those reported in previous studies \cite{Neri2020,Chicireanu2006}, and essentially unaffected by the presence of an overlapping  cloud of 10$^9$ Li atoms.
Moreover, while the optimum loading conditions strongly reduce the Cr MOT density \cite{Ciamei2022}, these do not limit the capture efficiency of the subsequent compressed-MOT (C-MOT) stage, operated at constant cooling light parameters: The strong increase in the MOT atom number thus directly turns into a significant density increase after the C-MOT, hence providing a substantial gain for the successive step of optical trap loading. 
I conclude this part emphasizing how this trick may be valuable to realize large MOTs of  for \textit{any} atomic species exhibiting strong light-assisted losses.

\subsubsection{\textit{Efficient loading of Cr-Li mixtures within an optical dipole trap}} \label{ODTLoad}
\begin{figure}
\centering
\includegraphics[width=1.\columnwidth]{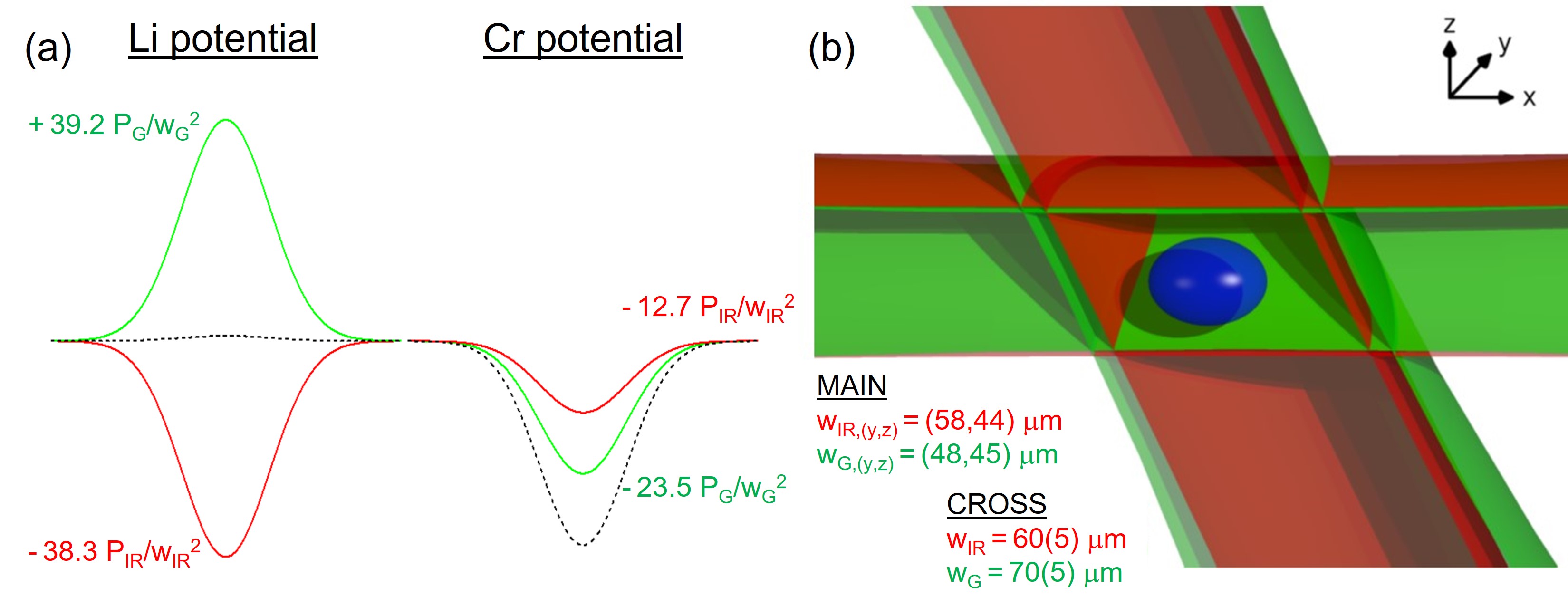}
\vspace*{-10pt}
\caption{
\textbf{(a)} Sketch of the optical potentials experienced by Li (left panel) and Cr (right panel) atoms confined in the BODT. The IR light (red curves) yields a trapping potential about 3.3 times deeper for Li than for Cr atoms, whereas the green beam (green curves) anti-confines lithium and tightly confines chromium. Trap depths in mK units are given in the sketch for powers ($P_{IR}$ and $P_{G}$) in Watts and beam waists ($w_{IR}$ and $w_{G}$) in micron. 
By adjusting the parameters of the two beams, here assumed to feature equal waists and power, one can obtain an overall BODT potential (black curves) more tightly confining for the Cr than for the Li component. 
\textbf{(b)} Schematic view of our BODT setup. Two overlapped IR and green beams, propagating in the horizontal plane along the $x$ direction, with 
waists indicated in the figure, provide the primary trapping potential for the atomic mixture, sketched in blue. A secondary bichromatic trap, realized by two additional overlapped IR and green circular beams, crosses the main BODT at an angle of about 15$^\circ$ from the vertical direction. Turning on the crossed BODT beam after the evaporation stage allows us to tune the densities of the two mixture components independently, while not modifying the trap depth.
}
\label{SketchODT}
\vspace*{-15pt}
\end{figure}

 As anticipated  above, two additional experimental challenges had to be overcome to produce Li-Cr mixtures in the ultracold regime: The difficulty of attaining efficient loading of the cold Cr cloud into a high-power IR optical dipole trap, and a chromium-to-lithium polarizability ratio for 1064 or 1070 nm IR lights 
 of only about 30$\%$, see Fig.~\ref{SketchODT}(a), not suited for efficient sympathetic cooling of Cr with Li. 
%
We could mitigate this second issue by superimposing to the IR trap a green beam at 532 nm, which tightly confines chromium while it anti-confines lithium, see green profiles in Fig.~\ref{SketchODT}(a). By tuning the relative power of the two lights of this bichromatic optical dipole trap (BODT), one can thus control the overall trap depth ratio for the two species, see black profiles in Fig.~\ref{SketchODT}(a).
Experimentally, the main BODT is realized by overlapping our IR trap, extensively discussed in Ref.~\cite{Simonelli2019} and based on a multimode fiber laser module from IPG Photonics (YLR-300) delivering up to 300 W, with a high-power laser at 532 nm (Sprout-G source by Lighthouse Photonics, nominally delivering up to 15 W).
Our main BODT beams, combined on a dichroic mirror, propagate in the horizontal $(x,y)$ plane, see Fig.~\ref{SketchODT}(b), and are focused onto the center of the Li-Cr MOT clouds.  

Opting for these two BODT wavelengths -- mainly motivated by the availability of high-power and relatively cheap laser sources -- allowed us to circumvent also the other technical issue, connected with detrimental light shifts that prevented the direct loading of Cr atoms from the MOT into an IR trap in previous experiments \cite{Chicireanu2007,Beaufils2008,Bismut2011}. Indeed, application of the IR light moves both ground and excited Cr levels -- connected by the main cooling transition -- to lower energy, with a shift for the excited state larger than the ground-state one. This makes that Cr atoms within an IR trap experience an effective  detuning of the MOT light that is effectively reduced -- and may eventually change sign -- causing strong heating and increased light-assisted losses of the chromium sample.
The choice of 532 nm light for our BODT setup naturally provided us an easy way to successfully circumvent this  issue: The key point is that the 532 nm light dramatically perturbs the Cr cooling transition, due to the presence of three strong atomic lines, connecting the excited level of the transition to higher-in-energy states, all centered around 533  nm \cite{Ciamei2022}. A relatively weak laser near 532 nm, blue-detuned from these lines by less than one nanometer, thus suffices to strongly shift the cooling transition towards higher frequencies, the effective MOT detuning experienced by atoms being in this case strongly moved \textit{away} from resonance.
The green light of our BODT could thus be efficiently exploited to (over-)compensate the detrimental effect of the IR trapping beam on the Cr (C-)MOT, realizing an effective `dark spot'.
Exploitation of such a trick, upon careful tuning of the green-to-IR power ratio, allowed us to substantially enhance the collection efficiency of chromium atoms (see Ref.~\cite{Ciamei2022} for more details): Under optimum conditions, up to 4 $\times$ 10$^6$ Cr atoms can be stored in the optical trap, at temperatures of about 250 $\mu$K, slightly lower than the typical C-MOT one. 
Finally, since the atom number transferred into the BODT is found to scale linearly with the MOT atom number itself \cite{Ciamei2022} -- with a constant 5.5$\%$ MOT-to-BODT transfer efficiency -- the `dark spot' strategy indeed appears to maintain light-assisted losses negligible up to the highest achievable densities, thereby making the ODT loading dynamics of chromium as simple as the one of lithium and other alkalis.

Besides being appealing also for single-species chromium experiments -- as it enables to collect a significant amount of $^{53}$Cr atoms -- this method turned out being especially advantageous in our mixture experiment:
Since the weak green beam does not perturb the Li loading dynamics, for this component we could indeed achieve MOT-to-BODT transfer efficiencies similar to those reported in Ref.~\cite{Burchianti2014} for the single-species case.
Integration of this method within our experimental cycle, combined with a short $D_1$ molasses phase on lithium  \cite{Burchianti2014} and hyperfine pumping stages, allows us to optically trap cold Li-Cr mixtures at about 250 $\mu$K, composed by 2$\times\!10^7$ $^6$Li atoms populating the two lowest Zeeman states $m_F\!=\!\pm\!1/2$ of the $F\!=\!1/2$ manifold, coexisting with about 2$\times\!10^6$ $^{53}$Cr atoms, asymmetrically distributed among the four lowest-lying Zeeman state of the $F\!=\!9/2$ hyperfine level. Specifically, without performing any Zeeman-selective optical pumping stage, about 55$\%$ of the Cr sample is found in the lowest Zeeman state, $m_F\!=\!-9/2$. The remaining Cr atoms are distributed among the three higher-lying levels, $m_F\!=\!-7/2$, $-5/2$ and $-3/2$, with relative populations of 25$\%$, 13$\%$ and 7$\%$, respectively 
\footnote{In the following, I denote the different Zeeman levels of both species with Li$|i\rangle$ and Cr$|i\rangle$ respectively, with $i\!=1,2,...$ labeling the atomic state starting from the lowest-energy one.}.
This represents the starting point of our evaporative and sympathetic cooling stages, lasting less than 5 s (see Fig.~\ref{ExpSketch}(a)), which are extensively discussed in  Ref.~\cite{Ciamei2022}, to which I refer the interested reader for details. 

\subsubsection{\textit{Degenerate Li-Cr Fermi mixtures and the `Feshbach cooling' mechanism}} \label{FCsec}
In the following, I  highlight another interesting trick -- unpublished thus far -- of potentially general interest, that was experimentally discovered by my colleagues S. Finelli and A. Ciamei. 
 This mechanism, denoted as `Feshbach cooling' by our team, allows us to substantially increase the Li-Cr thermalization rate at low temperatures, thereby substantially increasing the Cr sympathetic cooling efficiency, by relying on enhanced scattering rates close to a narrow inter-species FR.
Indeed, while the background Li-Cr scattering length, $a_{bg}\!\sim\!42a_0$  \cite{Ciamei2022B}, suffices to ensure inter-species thermalization from the initial temperature of about 250 $\mu$K down to a few $\mu$K, it is somewhat too small once lithium approaches the quantum-degenerate regime: From that point on, the chromium-to-lithium temperature ratio, $T_{Cr}/T_{Li}$, is found to progressively grow, as the  BODT trap depth is further decreased, see Fig. 7(d) of Ref.~\cite{Ciamei2022}.

Rather than circumventing this issue by significantly extending the evaporation ramps in this final stage, we opted to exploit the presence of various narrow $s$-wave Li-Cr Feshbach resonances, located at fields above 1400 G, where the Li$|1\rangle$-Li$|2\rangle$ scattering length is still large and negative, of about -2500 $a_0$ \cite{Zurn2013}. In particular, the Li$|1\rangle$-Cr$|1\rangle$ mixture possesses a $\sim$0.5 G-wide FR at 1414 G (with associated $R^*\!\sim$6000 $a_0$), and the Li$|2\rangle$-Cr$|1\rangle$ combination exhibits a resonance of similar character around 1461 G \cite{Ciamei2022B}, see also Sec.~\ref{FeshbachExp} below.
Both features are immune to two-body losses and, in spite of their relatively narrow character and high-field location, our magnetic-field stability, of only a few mG over hundreds of ms, allows us to finely tune the Li-Cr scattering length $a$ -- and thus the elastic scattering cross-section -- well above its background value.
\begin{figure}
\centering
\includegraphics[width=0.8\columnwidth]{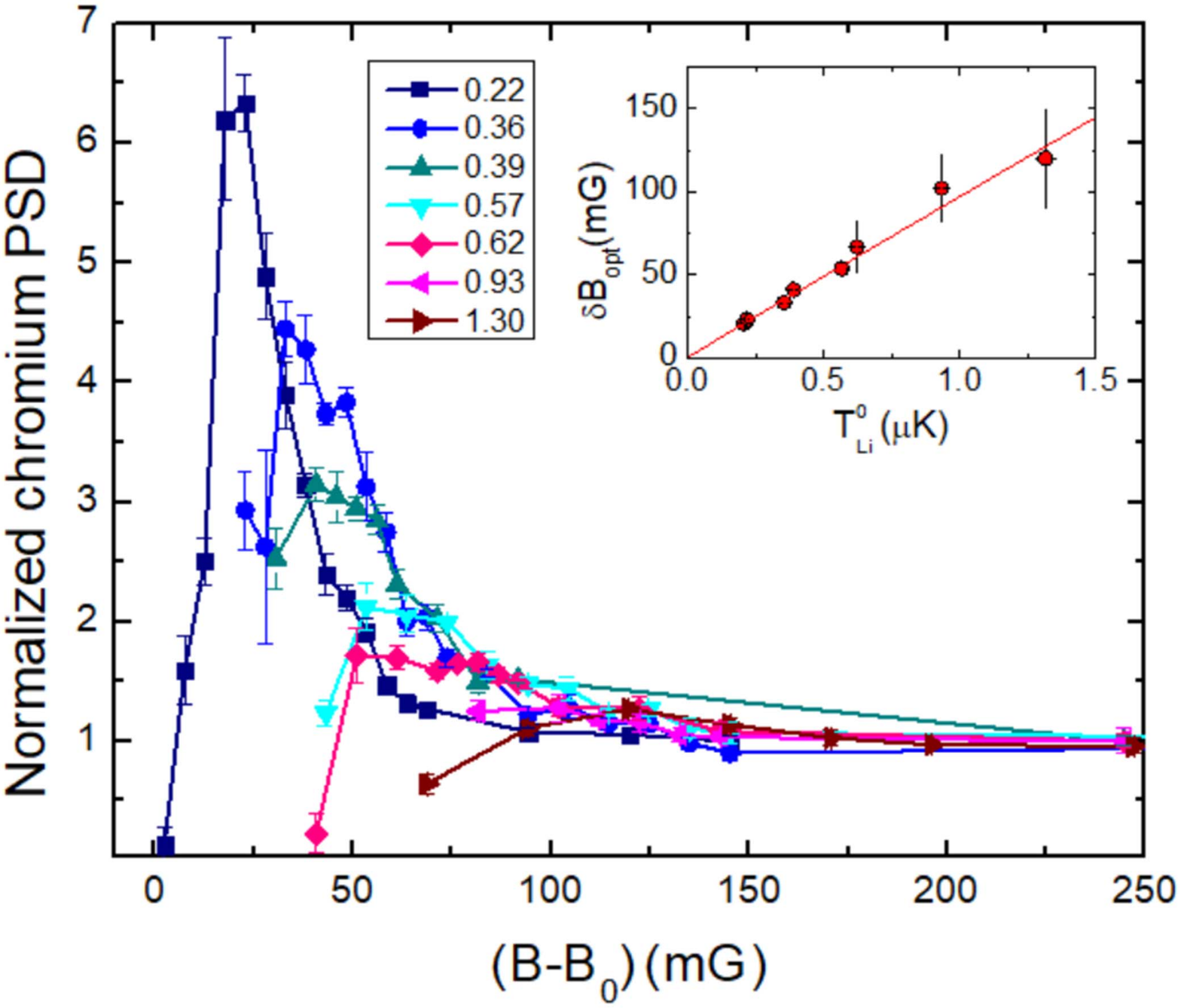}
\vspace*{-0pt}
\caption{
Main panel: Chromium phase-space density, normalized to its off-resonant value, is plotted as a function of the magnetic-field detuning $B-B_0$ from the Li$|1\rangle$-Li$|2\rangle$ FR centered at $B_0 \sim$1461 G \cite{Ciamei2022B}, where Cr is let thermalize with Li for 150 ms. 
Different datasets refer to different initial lithium temperatures, specified in the legend in $\mu$K units. For each measurement, the initial temperature mismatch was kept at $T_{Cr}^0/T_{Li}^0\sim$3. Inset: optimum detuning $\delta B_{opt}$, where the $PSD_{Cr}$ gain is found to be maximum, is plotted as a function of $T_{Li}^0$. Solid line is the best-fit to the data with a power-law function, yielding an exponent $\alpha$=0.99(3).
}
\label{FC1}
\vspace*{-15pt}
\end{figure}

In order to test such a possibility, about 4 s after the start of the evaporation, we intentionally established an initial temperature mismatch, $T_{Cr}^0/T_{Li}^0\sim$3 for various $T_{Li}^0$ values.
We then reduced the magnetic-field detuning from about +2 G above one of either Li-Cr resonances, down to a variable value $\leq$100 mG, correspondingly tuning the Li-Cr scattering length from $a\!\sim\!a_{bg}$ to $a\!\leq$-200 $a_0$, yet not significantly enhancing inelastic three-body loss processes.
We thus allowed the Li-Cr mixture to thermalize at the final field for a typical time of about 150 ms, after which we monitored atom number, temperature, and in-trap sizes of both species via absorption imaging. 
From these observables, we could also quantify the final gain in phase-space density (PSD) of chromium, $PSD_{Cr}\sim N_{Cr}/T_{Cr}^3$, in turn closely connected with an enhanced Li-Cr thermalization rate. 
Examples of $PSD_{Cr}$ data, normalized to their off-resonant value, are shown in the main panel of Fig.~\ref{FC1} as a function of the magnetic-field detuning, for various initial $T_{Li}^0$ values, see legend. These were extracted from Gaussian fits to the cloud imaged at long time-of-flight\footnote{For degenerate samples, $T_{Li}^0$ must be interpreted as a measure of the mean kinetic energy of the Li component.}.

These raw data show some interesting features, which nicely connect with the discussion made in Sec.~\ref{2B} about elastic scattering near narrow FRs, see Fig.~\ref{sigma_theo}.   
First, one can notice how for each initial temperature, there is an optimum detuning $\delta B_{opt}>0$, where the Cr PSD reaches a maximum value, which linearly decreases with the initial system temperature, see inset of Fig.~\ref{FC1}. 
Second, the lower is the sample temperature, the more sizable is the gain in PSD that can be achieved. Third, by tracing the time evolution of the chromium temperature at $\delta B_{opt}>0$ (not shown), we found that the extracted rate is inversely proportional to $\left\langle v_{rel}\right\rangle$, i.e. to the estimated Li-Cr mean relative velocity, obtained by averaging over the momentum distributions of the two components.
Finally, although not shown here, thermalization data taken at negative detunings, i.e. at $a>0$ values, do not exhibit any appreciable resonant enhancement at $\delta B\neq$0 as the one presented in Fig.~\ref{FC1}. 
Such a peculiar trend qualitatively matches the expectation from the two-body scattering theory discussed in Sec.~\ref{2B}. 
Indeed, as shown already in Fig.~\ref{sigma_theo}, for $R^*/|a|\!\gg$1 the cross-section becomes unitary-limited at finite $a\!<$0 values -- i.e. finite $\delta B\!>$0 in our case -- when $k\!=\!k_M\!\sim\!1/\sqrt{R^*|a|}$.
By equivalently expressing this relation in terms of collision energy $E\!=\!\hbar^2k^2/(2 m_r)$, and exploiting Eqs.~(\ref{aB2}) and (\ref{RsB}) for $a(B)$ and $R^*$, respectively, it is straightforward to obtain that collisions at energy $E$ are resonantly enhanced at the magnetic-field detuning $\delta B$ for which $E\!=\! \delta B \delta \mu$: Namely, when the collision energy equals the one of the (virtual) closed-channel molecule. Although this result does not account for thermal averaging over all Li-Cr collision energies, it qualitatively matches our observation of a linear dependence of $\delta B_{opt}$ upon the lithium temperature $T_{Li}^0$, see Fig.~\ref{FC1} inset.
Additionally, when the detuning is set at $\delta B_{opt}$ for a given momentum $\hbar k=m_r v_{rel}$, the corresponding scattering rate Eq.~(\ref{scatrate}) is expected to scale as $1/\tau \propto 1/v_{rel}$, again in agreement with our experimental observation.      
While a further theoretical analysis of the `Feshbach cooling' is currently in progress, it is important to stress that this phenomenon, exclusively determined by  \textit{elastic} scattering, is crucially relevant in our experimental routine to attain large, degenerate samples of lithium and chromium atoms, see Ref.~\cite{Ciamei2022}. Integration of the `Feshbach cooling' protocol within the evaporation stage performed in the main BODT, indeed allows us to produce up to 3.5$\times$10$^5$ Li$|1\rangle$ and 2.5$\times$10$^5$ Li$|2\rangle$ atoms at $T/T_{F,Li}\!\sim$0.25, coexisting with about 10$^5$ Cr$|1\rangle$ atoms at $T/T_{F,Cr}\!\sim$0.5 \cite{Ciamei2022}.
Besides being a key tool for the Li-Cr system, such a technique may be beneficial also for other ultracold mixtures -- Bose-Fermi, Bose-Bose, atom-ion, etc. -- and also single-species experiments, whenever narrow FRs (eventually not $s$-wave) are available. 
For instance, we are  aware that qualitatively similar observation has been obtained on Li-K (both Fermi-Bose and Fermi-Fermi) mixtures, as well as on homonuclear Dy Bose gases (Private communications by R. Grimm and G. Modugno's teams, respectively).

In conclusion of this section, I also mention an additional tool we devised in the experiment, represented by a second BODT crossed beam, see sketch in Fig.~\ref{SketchODT}(b): Being almost vertically oriented, this provides an additional confinement within the $(x,y)$ plane for both atomic species, while not significantly affecting their overall trap depth, hence their temperature.
Upon tuning the (absolute and relative) IR and green powers of this second BODT potential at the end of the evaporation stage, we can thus widely tune the density and degree of degeneracy of the two mixture components. This allows us to controllably pass from a highly-degenerate Li Fermi gas at $T/T_{F,Li}\!\leq$0.2 coexisting with a thermal Cr gas, to the opposite situation, where small Li samples are embedded within a chromium Fermi sea at $T/T_{F,Cr}\!\sim$0.2. 
Most importantly, over a sizable region of the crossed BODT parameter space, we can produce large and doubly-degenerate Li-Cr Fermi mixtures, where both component exhibit atom numbers of 10$^5$ or more, at reduced temperatures $T/T_{F}\!\sim$0.25, see Fig.~\ref{ExpSketch}(b). 
While referring the reader to Section V of Ref.~\cite{Ciamei2022} for further details, I remark here how this experimental tool offers a compelling opportunity for future investigation of the system phase diagram in the interaction-polarization plane, allowing us to smoothly range from the heavy-impurity limit -- crossing through the balanced case -- to the light-impurity one. 

\subsection{\textbf{$^6$Li-$^{53}$Cr scattering properties and Feshbach resonances}}\label{FeshbachExp}
 
In the following, I provide a brief overview of the scattering properties of $^6$Li-$^{53}$Cr Fermi mixtures which we  recently characterized \cite{Ciamei2022B} in a joined effort between our team -- that performed extensive loss-spectroscopy measurements -- and the theorist Prof. A. Simoni (Univ. Rennes), who could build, starting from experimental data, a full coupled-channel (CC) model for the Li-Cr system.
Experimentally, we investigated six different Li$|i\rangle$-Cr$|j\rangle$ spin-state combinations with $i \!=1,2$ and $j \!=1,2,3$, each being characterized by the total spin projection quantum number, $M_f\!=\!m_{f,Li}+m_{f,Cr}\!=\!-i+j-4$, thus spanning $-5\!\leq\!M_f\!\leq\!-2$. By scanning the magnetic field from 0 up to about 1500 G, we could pinpoint through loss spectroscopy more than 50 interspecies FRs.
In agreement with qualitative theoretical expectations already anticipated in Sec.~\ref{SecLiCrTheo}, the observed features are indeed arranged in complex but \textit{non-chaotic} patterns. 
Specifically, our measurements revealed qualitatively distinct features between $B$-field regions below and above 150 G, respectively: 
While few, sparse and narrow resonances characterize the $B\geq$150 G high-field region, more complex patterns  were found at lower fields, with strong loss peaks often arranged in doublet or triplet structures, see Fig. 1 of Ref.~\cite{Ciamei2022B}. 
This suggested that narrow and isolated $s$-wave features only occur in the high-field region, while the low-field spectra originate from strong FRs occurring in $l\!>0$ partial waves -- split by magnetic dipole-dipole interaction \cite{Ticknor2004,Cui2017} and possibly other couplings \cite{Zhu2019} -- as also supported by their sensitive dependence upon the system temperature \cite{Ciamei2022B}.

This intuition was indeed confirmed by the theoretical analysis carried out by A. Simoni based on full CC calculations, which account for: (i) The atomic hyperfine and Zeeman energies \cite{Childs1963,Arimondo1977}, defining the asymptotic collision thresholds; 
(ii) The strong and isotropic electrostatic interaction -- both $l$ and $M_f$ conserving -- represented by the \textit{sextet} $X ^6\Sigma^+$ and \textit{octet} $a ^8\Sigma^+$ potentials, parametrized by sextet $a_6$ and octet $a_8$ $s$-wave scattering lengths, respectively, as well as the dispersion coefficients $C_6$ and $C_8$; 
(iii) Weaker anisotropic couplings, originating from both long-range magnetic spin and short-range second-order spin-orbit interactions. These can couple different partial waves and hyperfine states with different $M_f$ or
$l$, provided that $\Delta l\!= 0,\pm 2$ and $M_f + m_l$ is conserved, $m_l$ being the projection of $l$ along the magnetic-field quantization axis \cite{Chin2010}.
Through least-square iterations by comparison with experimental data, A. Simoni could then optimize the initially unknown values of $a_6$, $a_8$, and of all other parameters entering the CC Hamiltonian  \cite{Ciamei2022B}. 
A global least-square fit -- able to reproduce all observed FR locations with remarkable accuracy -- provided the best-fit results $a_6\!=\!15.46(15)~a_0$, $a_8\!=\!41.48(2)~a_0$, $C_6\!=\!922(6)~a.u.$, and $C_8\!=9.8(5)\,10^4~a.u.$, with errors denoting one standard deviation obtained from the fit covariance matrix. 

The Li-Cr scattering properties obtained from our optimized CC model can be summarized as follows:
First, the Li-Cr mixture exhibits a background $s$-wave scattering length, which is almost everywhere close to the octet $a_8$ value, of about 42 $a_0$, only slightly lower than the Li-K one, of about 65 $a_0$ \cite{Naik2011}. 
The low-field  spectral region is entirely dominated by $p$-wave FRs, featuring  $m_l$ splittings much larger than those found in alkali systems \cite{Ticknor2004,Cui2017,Zhu2019}, owing to the increased role of spin-spin dipole coupling in Li-Cr mixtures, and to the coincidentally small relative magnetic moment of the molecular states involved.
Particularly interesting in this respect is the presence, in the Li$|2\rangle$-Cr$|1\rangle$ mixture, of a strong $p$-wave FR with $m_l\!=\!-1$, centered around 24 G and essentially immune to two-body losses. This feature,  which is intrinsically \textit{chiral} in nature, could serve for future many-body surveys of $p$-wave resonant Fermi mixtures \cite{Gurarie2005} in yet unexplored regimes. 
Indeed, since the $B$-field splitting between different FRs with different $m_l$ values exceeds several Gauss, one has the rather unique possibility of resonantly enhance \textit{only} scattering of colliding Li-Cr pairs that exhibit, say, a $p_x\!-\!ip_y$ orbital symmetry. Hence, in contrast with the case  investigated so far in homonuclear Fermi gases \cite{Regal2003B,Zhang2003}, a gas of $N$ LiCr $p$-wave dimers at such FR should carry a macroscopic angular momentum of order $N \hbar$, possibly realizing, in the degenerate regime, a \textit{superfluid orbital ferromagnet} \cite{Ho2005}.    

 Also several $s$-wave FRs are found within various Li-Cr spin-combinations. 
Owing to the relatively small values of $a_6$ and $a_8$, similarly to the Li-K case \cite{Wille2008,Naik2011}, these features are generally narrow in nature, since the FR character -- as for bi-alkalis \cite{Chin2010} -- is determined by the difference between sextet and octet scattering lengths, the larger being this difference, the wider the corresponding features \cite{Chin2010}. 
Specifically, our model connects all FRs observed above 1400 G to $l_r\!=\!0$ molecular levels of $X ^6\Sigma^+$ potentials, all featuring magnetic-field widths $\Delta B\!\sim$0.5~G and differential magnetic moments $\delta \mu\!=$2 $\mu_B$, yielding effective-range parameters of $R^*\!\sim$6000~$a_0$, in perfect agreement with Eq.~(\ref{RsB}). 
Most importantly, one of these features occurs around 1414 G within the absolute Li$|1\rangle$-Cr$|1\rangle$ ground state, thus completely immune to two-body inelastic losses, and a second one is found at 1461 G for the Li$|2\rangle$-Cr$|1\rangle$ mixture (see Fig.~\ref{CCvsZR}), also exhibiting negligible dipolar relaxation rates. 
It is also interesting to remark a peculiarity of this family of FRs, i.e. the nearly-coincident location of features occurring in Li$|i\rangle$-Cr$|j\rangle$ and Li$|i\rangle$-Cr$|j+1\rangle$ combinations: For instance, the Li$|1\rangle$-Cr$|2\rangle$ FR occurs around 1418 G, i.e. only 4 G above the ground-state one, and similar $B$-field gaps are found for higher-lying mixtures involving the Li$|2\rangle$ state \cite{Ciamei2022B}.  
 
The fact that $^6$Li-$^{53}$Cr exhibits $s$-wave FRs immune to two-body losses represents a fundamental advantage, with respect to the Li-K case: Combined with an increased Li-Cr mass ratio, which lays \textit{above} the critical one for KM trimers to appear, this makes the lithium-chromium system extremely promising for future few- and many-body studies, in spite of the narrow character of the available FRs. 
Although in our setup we could already achieve a very good magnetic-field stability -- better than 5 mG peak-to-peak around 1400 G over $>$100 ms -- stable LiCr$_2$ trimers will likely not be observable in dilute three-dimensional samples, owing to the ultra-narrow $B$-field region corresponding to $R^*/a\leq$0.03, see Fig.~\ref{Cr2Li}. 
Nonetheless, the associated Cr-LiCr atom-dimer $p$-wave resonance will still be fully exploitable, with resonant $p$-wave Cr-dimer attraction being expected to overcome the concurrent $s$-wave repulsion over a few-tens-of-mG wide region below the FR pole, as for the Li-K case, see Ref.~\cite{Jag2014}. 
Additionally, stable LiCr$_2$ trimers (and possibly also LiCr$_3$ tetramers) should become experimentally detectable in a moderate quasi-2D environment \cite{Levinsen2009}, or in presence of a degenerate Cr Fermi sea \cite{Mathy2011}. 

First studies, and yet ongoing investigation, of inelastic three-body loss rates appear promising in view of future many-body surveys, as they point to lifetimes of Li-Cr Fermi mixtures exceeding a few tens of ms under resonant conditions \cite{Ciamei2022B}, in spite of the sizable $R^*$ value of the available FRs.   
As such, the narrow character of the $s$-wave features, although being expected to cause modification of the system  phase diagram \cite{Radzihovsky2010}, does not appear to prevent the exploration of Li-Cr Fermi mixtures under strong (attractive and repulsive) interactions. 
In this respect, it is worth noticing how the high-field location of these FRs -- indeed quite scary at a first glance -- turns out being extremely advantageous for the implementation of radio-frequency spectroscopy schemes. 
Owing to the fact that, at such fields, both $^6$Li and $^{53}$Cr atoms are deep in the Paschen–Back regime, the differential magnetic moment between nearby Zeeman levels is drastically reduced. 
For instance, at 1400 G, this amounts to only 0.77 kHz/G for the Li$|1\rangle\!\rightarrow$Li$|2\rangle$ transition (around 80 MHz), and to less than 6 kHz/G for the Cr$|1\rangle\!\rightarrow$Cr$|2\rangle$ one (around 240 MHz), hence making them extremely \textit{insensitive} to residual $B$-field fluctuations. With our field stability, this allows us to implement exceedingly long spectroscopy pulses of several tens, or even hundreds, of ms. 
As shown by the proof-of-principle measurement presented in Fig.~\ref{ExpSketch}(c) -- intentionally performed on Li atoms embedded in a dilute, thermal gas of  chromium -- this possibility results in an extreme ($\sim$1 Hz) sensitivity to interaction-induced shifts, when investigating subtle few-body phenomena, heavy and light impurity problems, as well as paired states or induced-interaction effects within Li-Cr fermionic mixtures.
Note that a similar feature in principle occurs for bound-to-bound transitions of the kind Li$|i\rangle$Cr$|j\rangle \rightarrow$Li$|i\rangle$Cr$|j+1\rangle$, connecting weakly-bound molecular states associated with nearly-coincident FRs at high field, and off-resonant with respect to the corresponding atomic transitions. This  appears as a very appealing opportunity for the investigation of the quasi-particle properties of LiCr dimers, dressed by their interaction with a surrounding bath of either heavy or light Fermi gases.

Finally, recent  (yet unpublished) characterization of LiCr dimer magneto-association at both Li$|1\rangle$-Cr$|1\rangle$ and Li$|2\rangle$-Cr$|1\rangle$ FRs, points to very high atom-to-molecule conversion efficiencies exceeding 50$\%$, with up to 6$\times$10$^4$ dimers being formed, at phase-space densities well above 0.1, see Fig.~\ref{ExpSketch}(d). 
Moreover, we already have experimental evidence for the increased dimer stability against inelastic collisions close to the FR pole: While the overall molecule lifetime so far is limited to about 15 ms near the FR pole, this is mainly ascribable to photo-excitation processes, induced by our BODT  lights, similarly to what recently reported for KDy dimers in Ref.~\cite{Soave2023}, see R. Grimm's contribution to these Proceedings. 
In the absence of this one-body loss contribution -- which can be experimentally solved by optimizing the wavelength of the trapping laser(s) -- our preliminary measurements point to a LiCr molecule lifetime, mainly limited by Li-dimer collisions in agreement with Fig.~\ref{etaAD} expectation, exceeding several tens of ms near resonance at particle densities of about  10$^{12}$ $cm^{-3}$. 
This makes the Li-Cr system a very promising candidate to produce large and long-lived samples of bosonic Feshbach dimers at high phase-space densities and, possibly, molecular Bose-Einstein condensates. 
In turn, this would lay the ground to the exploration of resonant superfluidity with a mass-imbalanced Fermi mixture, and it would also represent an optimal starting point to realize,  through optical transfer schemes towards deeply-bound levels, degenerate Bose gases of paramagnetic polar molecules \cite{Zaremba2022}.
 
\section{Conclusive remarks} \label{Conclusions}

In this Lecture, I summarized the basic theoretical tools -- more extensively presented in Refs.~\cite{Levinsen2011,Petrov2012} -- that should allow also a non-expert reader to understand the rich phenomenology associated with the few-body physics of mass-imbalanced Fermi mixtures, and its possible implication at the many-body level.
Building on such general framework, I discussed theoretical predictions and expected peculiarities of $^6$Li-$^{53}$Cr Fermi mixtures, which are enabled by the specific mass ratio between the two mixture components and -- as LiCr ground-state molecules are concerned-- by the non-alkaline nature of the chromium transition-metal element.
Finally, I highlighted the main features of this novel lithium-chromium system, that we could already gain from recent and ongoing experimental activities performed in our lab. While many more details can be already found in Refs.~\cite{Simonelli2019,Neri2020, Ciamei2022B,Ciamei2022}, some of the topics covered by these notes will be subject of forthcoming publications.

To conclude, the future of experimental research on lithium-chromium mixtures appears bright and promising: As  several non-trivial conceptual and technical issues of such a novel ultracold system have been already overcome in our lab, the realization of large and deeply-degenerate $^6$Li-$^{53}$Cr samples, and the detailed knowledge of their collisional properties, provide an important starting point for further exciting studies, encompassing novel types of few-body phenomena and impurity problems, non-trivial many-body regimes of highly-correlated fermionic matter, and realization of ultracold quantum gases of paramagnetic polar molecules.

\acknowledgments
I first thank D. Petrov and R. Grimm for having introduced me to the topics discussed in these notes, allowing me to understand several theoretical and experimental aspects of strongly-interacting Fermi mixtures. 
Special thanks to all people of the PoLiChroM lab, and in particular to  A. Ciamei, S. Finelli, E. Neri and A. Trenkwalder,  whose work over the years allowed to turn a set of ideas into reality. 
I also acknowledge G. Bruun,  W. Ketterle, B. Laburthe-Tolra, J. Levinsen, P. Massignan, M. Parish, T. Pfau, R. Schmidt, A. Simoni, M. Tomza, L. Vernac, and all  members of the Quantum Gases Group in Florence for stimulating discussions and fruitful collaboration. 
Financial support is acknowledged from  the European Research Council under Grant no. 637738 (PoLiChroM), from the Italian MIUR through the FARE Grant no. R168HMHFYM (P-HeLiCS), and from the EU H2020 Marie Skłodowska-Curie grant agreement no. 894442 (CriLiN).





%

\end{document}